\documentclass[sigconf, screen]{acmart}

\usepackage{multirow}
\usepackage{colortbl}
\usepackage{xspace}
\usepackage{graphicx}
\usepackage{xcolor}
\usepackage{subcaption}
\usepackage{soul}
\usepackage{fontawesome5}
\usepackage{tcolorbox}
\usepackage{color}
\usepackage{rotating}
\usepackage{tabularray}
\usepackage{graphicx} 
\usepackage{makecell}
\usepackage{pifont}
\usepackage{hyperref}
\usepackage{float} 
\usepackage{tabularx}
\usepackage{enumitem}
\usepackage{hhline}
\usepackage{algpseudocode}
\usepackage{listings}
\usepackage{xcolor}
\usepackage{xspace}
\usepackage{algorithm}
\usepackage{amsmath}

\definecolor{rqblue}{RGB}{100,160,220}
\definecolor{rqbluebg}{RGB}{240,248,255}
\definecolor{rqgreen}{RGB}{120,190,120}
\definecolor{rqgreenbg}{RGB}{242,250,242}
\definecolor{rqorange}{RGB}{240,170,100}
\definecolor{rqorangebg}{RGB}{255,248,240}
\definecolor{rqlightyellow}{RGB}{255,205,179}
\definecolor{rqlightyellowbg}{RGB}{255,248,240}
\definecolor{rqlightblue}{RGB}{131,202,234}
\definecolor{rqlightbluebg}{RGB}{249,252,254}
\definecolor{rqlightgreen}{RGB}{217,241,208}
\definecolor{rqlightgreenbg}{RGB}{251,254,250}
\definecolor{rqpurple}{RGB}{170,140,200}
\definecolor{rqpurplebg}{RGB}{248,245,252}
\definecolor{rqred}{RGB}{220,100,100}
\definecolor{rqredbg}{RGB}{255,245,245}

\newcommand{\answer}[4][]{%
  \begin{tcolorbox}[
    colframe=rq#3,
    colback=rq#3bg,
    coltitle=black,
    fonttitle=\bfseries,
    fontupper=\color{black},
    boxrule=0.8pt,
    arc=3pt,
    left=5pt,right=5pt,top=5pt,bottom=5pt,
    title={\if\relax\detokenize{#1}\relax Recommendation~#2\else #1~#2\fi}
  ]
  #4
  \end{tcolorbox}
}

\AtBeginDocument{%
  }

\setcopyright{acmlicensed}
\copyrightyear{2018}
\acmYear{2018}
\acmDOI{XXXXXXX.XXXXXXX}

\acmConference[Conference acronym 'XX]{Make sure to enter the correct
  conference title from your rights confirmation emai}{June 03--05,
  2018}{Woodstock, NY}
\acmISBN{978-1-4503-XXXX-X/18/06}

\newif\ifcomment

\ifcomment
  \newcommand{\gao}[1]{\textcolor[rgb]{0.6,0,0.2}{Jie: #1}}
    
  \newcommand{\shunyi}[1]{\textcolor[rgb]{0.012, 0.659, 0.620}{Shunyi: #1}}
  \newcommand{\zoey}[1]{\textcolor[rgb]{0.01, 0.659, 0.20}{Zoey: #1}}
  \newcommand{\ziang}[1]{\textcolor[rgb]{0.012, 0.659, 0.620}{Ziang: #1}}
  \newcommand{\alok}[1]{\textcolor[rgb]{0.012, 0.659, 0.620}{Alok: #1}}
  \newcommand{\cmh}[1]{\textcolor{blue}{Chien-Ming: #1}}

\else
  \newcommand{\gao}[1]{}
  \newcommand{\shunyi}[1]{}
  \newcommand{\zoey}[1]{}
  \newcommand{\ziang}[1]{}
  \newcommand{\alok}[1]{}
  \newcommand{\cmh}[1]{}
    
\fi

\newcommand{\mindcoder}{\textit{MindCoder}\xspace}

\newcolumntype{L}[1]{>{\arraybackslash}p{#1}}

\lstdefinelanguage{json}{
    morestring=[b]",
    morecomment=[l]{//},
    morecomment=[l]{\#},
    literate=
      *{0}{{{\color{black}0}}}{1}
       {1}{{{\color{black}1}}}{1}
       {2}{{{\color{black}2}}}{1}
       {3}{{{\color{black}3}}}{1}
       {4}{{{\color{black}4}}}{1}
       {5}{{{\color{black}5}}}{1}
       {6}{{{\color{black}6}}}{1}
       {7}{{{\color{black}7}}}{1}
       {8}{{{\color{black}8}}}{1}
       {9}{{{\color{black}9}}}{1}
       {:}{{{\color{black}{:}}}}{1}
       {,}{{{\color{black}{,}}}}{1}
       {[}{{{\color{black}{[}}}}{1}
       {]}{{{\color{black}{]}}}}{1}
       {\{}{{{\color{black}{\{}}}}{1}
       {\}}{{{\color{black}{\}}}}}{1}
}
\begin{document}


\title{Efficiency with Rigor! A Trustworthy LLM-powered Workflow for Qualitative Data Analysis}


\author{Jie Gao}
\authornote{This work was partially done when the author was a postdoctoral researcher at Singapore-MIT Alliance for Research and Technology.}
\affiliation{%
  \institution{Johns Hopkins University}
  \city{Baltimore}
  \state{MD}
  \country{USA}
}
\email{jgao77@jh.edu}

\author{Zhiyao Shu}
\affiliation{%
 \institution{University of California, Berkeley}
 \city{Berkeley}
 \state{CA}
 \country{USA}
 }
\email{yaoshu0326@berkeley.edu}

\author{Shun Yi Yeo}
\affiliation{%
 \institution{Singapore University of Technology and Design}
 \country{Singapore}
 }
\email{yeoshunyi.sutd@gmail.com}

\author{Alok Prakash}
\affiliation{%
 \institution{Singapore-MIT Alliance for Research and Technology}
 \country{Singapore}
 }
\email{alok.prakash@smart.mit.edu}

\author{Chien-Ming Huang}
\affiliation{%
 \institution{Johns Hopkins University}
 \city{Baltimore}
  \state{MD}
  \country{USA}
 }
\email{cmhuang@cs.jhu.edu}

\author{Mark Dredze}
\affiliation{%
 \institution{Johns Hopkins University}
 \city{Baltimore}
  \state{MD}
  \country{USA}
 }
\email{mdredze@cs.jhu.edu}

\author{Ziang Xiao}
\affiliation{%
 \institution{Johns Hopkins University}
 \city{Baltimore}
  \state{MD}
  \country{USA}
 }
\email{ziang.xiao@jhu.edu}

\renewcommand{\shortauthors}{Trovato et al.}

\begin{abstract}
Qualitative data analysis (QDA) emphasizes trustworthiness, requiring sustained human engagement and reflexivity. Recently, large language models (LLMs) have been applied in QDA to improve efficiency. However, their use raises concerns about unvalidated automation and displaced sensemaking, which can undermine trustworthiness. To address these issues, we employed two strategies: transparency and human involvement. Through a literature review and formative interviews, we identified six design requirements for transparent automation and meaningful human involvement. Guided by these requirements, we developed MindCoder, an LLM-powered workflow that delegates mechanical tasks, such as grouping and validation, to the system, while enabling humans to conduct meaningful interpretation. MindCoder also maintains comprehensive logs of users’ step-by-step interactions to ensure transparency and support trustworthy results. In an evaluation with 12 users and two external evaluators, MindCoder supported active interpretation, offered flexible control, and produced more trustworthy codebooks. We further discuss design implications for building human-AI collaborative QDA workflows.

\end{abstract}

\ccsdesc[500]{Human-centered computing~Interactive systems and tools}

\keywords{Reasoning, Chain-of-Thought Prompting, Prompting Techniques, Qualitative Coding, Large Language Models, Code-to-theory Model}

\begin{teaserfigure}
\vspace{-10pt}
  \includegraphics[width=\textwidth]{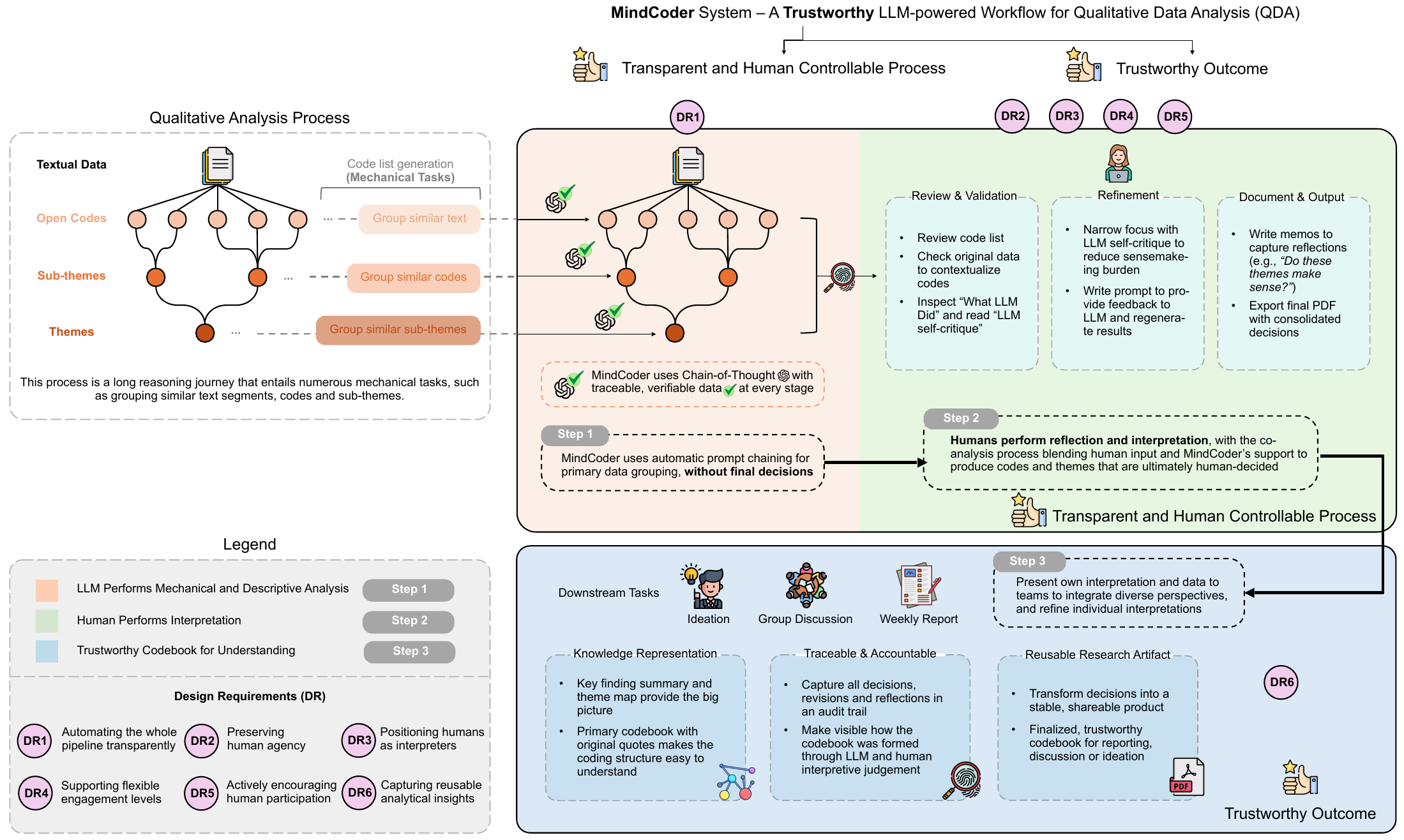}
  \vspace{-15pt}
  \caption{\href{https://mindcoder.ai/}{MindCoder}, A Trustworthy LLM-powered Workflow for Qualitative Data Analysis}
  \label{fig:teaser}
\end{teaserfigure}



\maketitle

\section{Introduction}

The development of Qualitative Data Analysis (QDA) methods in the past decades \cite{flick2013sage}, including Grounded Theory \cite{bryant2010sage} and Thematic Analysis \cite{braun2006using}, has aimed to establish rigorous approaches for systematically analyzing qualitative data \cite{hecker2025transparency, nowell2017thematic, lincoln1985naturalistic}. A central goal of these methods is establishing trustworthiness \cite{amin2020establishing}, defined as whether findings are \textit{``worth paying attention to, worth taking account of''} \cite{lincoln1985naturalistic}. Trustworthiness in QDA encompasses four components: credibility (truth of findings), transferability (applicability across contexts), dependability (replicability), and confirmability (minimization of bias). Achieving such rigor demands intensive human involvement through sustained engagement with data, researcher reflexivity, detailed audit trails, and collaborative verification processes \cite{nowell2017thematic, ahmed2024pillars}.

Meanwhile, efficiency is another critical goal in QDA \cite{baba2011nvivo, weitzman1995computer}. An efficient QDA allows researchers to handle larger datasets, reduce mechanical tasks, and devote more of their time to interpreting the data and developing theory. To this end, researchers have relied on computer-assisted qualitative data analysis software (CAQDAS) \cite{banner2009computer}, such as NVivo \cite{baba2011nvivo}, MaxQDA \cite{gizzi2021practice}, and Atlas.ti \cite{smit2002atlas}. More recently, advances in large language models (LLMs) have transformed the field: their ability to rapidly process, interpret, and generate text analyses has driven adoption in domains ranging from human–computer interaction \cite{marianne2024llm, quere2025stateofLLM} and software engineering \cite{lecca2024applications, rasheed2024largelanguagemodelsserve} to education \cite{barany2024chatgpt} and social science \cite{paoli2024performing, liu2023voices, shen2023shaping}. Researchers are increasingly open to integrating LLMs throughout the research pipeline \cite{schroeder2025largelanguagemodelsqualitative}, and most CAQDAS platforms now include LLM-powered features\footnote{Atlas.ti \url{https://atlasti.com/ai-coding-powered-by-openai}}\footnote{MaxQDA \url{https://help.maxqda.com/en/support/solutions/folders/80000724281}}.

However, as LLMs enter the practice of QDA, they promise efficiency but raise concerns within the scientific community for their validity \cite{wiebe2025qualitative, schroeder2025largelanguagemodelsqualitative}. The opacity and stochasticity of LLMs make it challenging to verify how conclusions are reached, undermining the methodological rigor of trustworthy QDA \cite{nyaaba2025optimizing}. Without rigorous human validation, LLM-assisted analysis risks producing ``sloppy'' research that seems high-quality but lacks trustworthiness \cite{LissackMeagher2024SloppyScience}. Most critically, automation risks eliminating the human interpretation that gives qualitative research its depth \cite{schroeder2025largelanguagemodelsqualitative}. Researchers may accept LLM outputs at face value rather than engaging deeply with their data. 
Together, these critiques raise a central challenge: \textbf{How can researchers harness LLMs' efficiency while ensuring the methodological rigor for trustworthy QDA?}

The research community has proposed various ways to strengthen trustworthiness in LLM use. For example, in deductive coding, Oksanen et al. \cite{oksanen2025llmcode} recommend researcher–AI alignment as an indicator of quality, suggesting that researchers report alignment metrics and examine worst-case performance to evaluate human–LLM alignment, thereby ensuring robust insights. For inductive, in LLM-led stepwise thematic coding, Nyaaba et al. \cite{nyaaba2025optimizing} highlight data traceability, requiring models to output exact sentences from source texts along with page numbers to support verification. In a human–LLM workflow, Wiebe et al. \cite{wiebe2025qualitative} emphasize accountability by documenting human coding roles and supporting human review of intermediate steps. Together, these works represent early steps toward establishing trustworthiness. However, what remains underexplored is how abstract concepts of trustworthiness can be translated into concrete practices, particularly regarding transparency and human involvement, that are both acceptable to analysts and meaningful in application.

\begin{figure*}[!t]
    \centering
    \includegraphics[width=\linewidth]{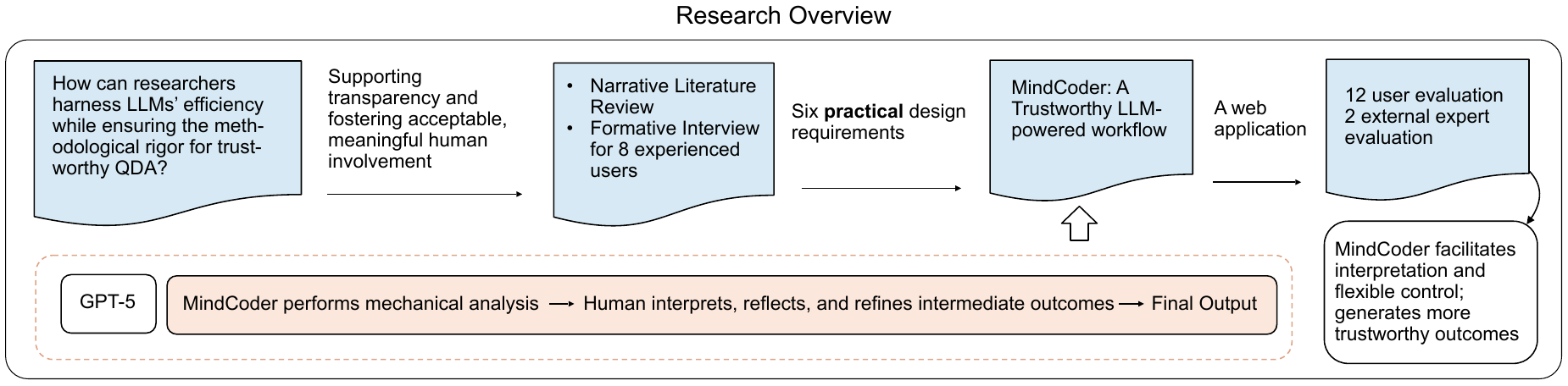}
    \caption{An overview of our motivation and method.}
    \label{fig:research-overview}
\end{figure*}

To bridge this gap, we conducted a narrative literature review and a formative study with 8 experienced QDA practitioners, which led us to identify 6 key design requirements (DR) for building the trustworthy LLM-powered QDA workflow: automating the entire pipeline through transparent reasoning chain (DR1), preserving analyst agency (DR2), positioning humans as interpreters rather than validators (DR3), supporting flexible engagement levels (DR4), actively encouraging participation (DR5), and capturing reusable analytical insights (DR6). These DRs provide practical guidelines for ensuring that humans drive interpretation while benefiting from LLM automation.

Guided by these DRs, we developed MindCoder, a human-LLM collaborative QDA workflow (Figure \ref{fig:teaser}, \ref{fig:workflow}) including three steps. First, MindCoder conducts preliminary coding using a transparent reasoning chain: it groups text into codes and (sub)themes based on similarity, and then validates them against the original data. The intermediate steps and outcomes are then exposed to analysts for review. Second, analysts engage with these intermediate artifacts to interpret and reflect, supported by reflective nudges such as LLM rationales, LLM self-critic and memo writing. Finally, MindCoder automatically consolidates the results into a Trustworthy Codebook, including user analysis trajectory, primary codebook, theme map, and original data, supporting downstream reuse such as group discussions.


To evaluate MindCoder, we conducted a within-subject study with 12 participants, comparing its use to one of the state of the art CAQDAS software (Atlas.ti AI Coding powered by OpenAI model). The participants’ analysis outcomes were further evaluated by two external expert annotators. Our findings indicate that MindCoder (1) fosters more active reflection and interpretation, (2) accommodates users’ needs for flexible control across different analytical tasks, and (3) generates outcomes that downstream users perceive as significantly more trustworthy. Moreover, our study highlights that trustworthiness is defined differently across stakeholder groups. For analyst users who conduct the analysis, it is assessed intuitively through proxies such as the ability to revise outputs and traceability to the original text. For downstream users, trustworthiness is evaluated using explicit metrics such as credibility, transparency, and verifiability of the analysis outcomes. For clarity, our research method is summarized in Figure \ref{fig:research-overview}.

Our contributions are threefold:

\begin{itemize}
\item We present the first systematic investigation of trustworthiness in LLM-powered QDA, translating classical validity criteria and expert insights into actionable design requirements that can guide the development of future LLM-powered QDA systems.
\item We translated these requirements into concrete design features implemented in MindCoder, a web application for trustworthy LLM-powered QDA.
\item We evaluated MindCoder through a within-subject study where 12 users analyzed qualitative data, with their analyses results independently assessed by two QDA experts. Results demonstrated that our design supports transparency and meaningful human involvement, producing trustworthy QDA outcomes.
\end{itemize}
\section{Related Work}
\subsection{Trustworthiness and Strategies in QDA}

In QDA, researchers must demonstrate that their results are "meaningful and useful"—that is, convincing to audiences (including themselves) that the findings are \textit{"worth paying attention to, worth taking account of"} \cite{lincoln1985naturalistic, ahmed2024pillars, amin2020establishing}. The framework of trustworthiness, proposed by Lincoln and Guba \cite{lincoln1985naturalistic}, provides four criteria for this evaluation: credibility, transferability, dependability, and confirmability.  
To build credibility (truth of findings), researchers may engage in the data to understand perspectives in the data, reflect on their own biases and preconceptions, and use triangulation across data sources or methods to cross-verify findings.  
To build transferability (applicability to other contexts), researchers provide thick descriptions, including rich narratives of the setting and context, to enable others to judge whether the findings are relevant or comparable to different situations.
To build dependability (replicability of findings) and confirmability (minimization of bias), researchers provide transparent documentation of procedures and decisions, maintaining an audit trail that records how the study was conducted, why particular choices were made, and how data were analyzed, allowing others or alternative perspectives to validate and confirm the accuracy of the findings.

In particular, \textit{transparency}, \textit{prolonged human involvement}, and the \textit{audit trail} are critical strategies to ensure trustworthiness \cite{amin2020establishing, ahmed2024pillars}.
For \textit{transparency}, researchers highlight the importance of recording justifications, intermediate analytical steps, and opportunities for human review and intervention \cite{Ailyze2025_RigorTransparency, DeCarlo2021_QualityRigor, ahmed2024pillars}.  
For \textit{prolonged human involvement}, researchers should apply critical interpretive judgments, enabling nuanced insights into participants’ experiences, behaviors, and beliefs \cite{ahmed2024pillars}. 
Finally, for the \textit{audit trail}, researchers should record key decisions throughout the process, allowing readers to trace the logic of the study and evaluate the reliability of its findings \cite{carcary2009research}. 
Taken together, these strategies collectively strengthen the \textit{trustworthiness} of qualitative research.  

In particular, establishing trustworthiness in QDA requires substantial human involvement, from reviewing intermediate steps to making critical interpretations. In this paper, we focus on transparency and human involvement (treating audit trails as one form of transparency) and examine how these high-level concepts can be translated into designs requirements that enable meaningful human–LLM collaboration for QDA in practice.

\subsection{LLM-powered QDA Workflow}
Recently, LLMs have been applied in QDA to support the efficient QDA, attracting researchers from various disciplinaries, including human–computer interaction \cite{marianne2024llm, quere2025stateofLLM}, software engineering \cite{lecca2024applications, rasheed2024largelanguagemodelsserve}, education \cite{barany2024chatgpt}, social science \cite{paoli2024performing, liu2023voices, shen2023shaping}, and other disciplines \cite{perkins2024generativeaitoolsacademic}. Early explorations have applied LLMs to classical GT \cite{sinha2024role} and TA \cite{wiebe2025qualitative}, reporting that, with tailored prompting strategies and customizations, LLMs can potentially produce promising results \cite{hou2024prompt, de2025reflections, katz2024thematic, sun2024eliciting, dai2023llm}. Specifically, existing LLM-based approaches for QDA typically include \textit{LLM-led analysis} and \textit{human-led analysis}.

\subsubsection{LLM-led analysis}
\textit{LLM-led analysis} involves minimal human involvement, such as prompt engineering \cite{zhang2025harnessingredesigning} or post-hoc refinement \cite{barany2024chatgpt}. These methods often prioritize efficiency but raises quality concerns, because LLMs are prone to hallucinations \cite{bano2024large}, embedded biases \cite{schroeder2025largelanguagemodelsqualitative}, and opaque “black-box” processes \cite{nyaaba2025optimizing}. Such opacity can exclude humans from critical stages of coding and categorization, fostering unwarranted trust in outputs and heightening the risk of overreliance, particularly in contexts where efficiency is prioritized.  

\subsubsection{Human-led analysis} In contrast, \textit{human-led analysis} involves extensive human involvement but often limits the efficiency gains of LLMs. In this approach, LLMs are mainly used for code or theme suggestions while humans retain responsibility for interpretation and analysis \cite{overney2024sensemate, palea2024annota, collabcoder2024gao}. 
Moreover, traditional CAQDAS, such as NVivo, MAXQDA, and ATLAS.ti, simulate the manual process in which coding was once performed entirely by hand with pen and paper \cite{baba2011nvivo, ngulube2023improving}, provides transparent and traceable coding processes \cite{bryda2023qualitative}, which could help researchers build trustworthy QDA. Recently, they have begun integrating LLMs\footnote{Atlas.ti: \url{https://atlasti.com/ai-coding-powered-by-openai}}%
\footnote{MaxQDA: \url{https://help.maxqda.com/en/support/solutions/folders/80000724281}}, enabling users to benefit from LLM capabilities while leveraging the established features of these tools. 
However, both of them offers limited opportunity for code generation across the full pipeline — from codes to subthemes and from subthemes to higher-level categories. These approaches still rely heavily on humans for basic verification and offer little optimization for meaningful or efficient involvement.

Yet most users seek a balance: conducting QDA as efficiently as possible while ensuring trustworthy findings. In this paper, we examine whether transparent workflows with intermediate human control and carefully designed involvement can balance efficiency and trustworthiness by managing trade-offs between two workflow modes.


\subsection{Human Involvement in QDA}

Designing an LLM-powered workflow requires a clear understanding of human roles and responsibilities, as this directly influences how tasks should be designed. In QDA context, analysis typically happens in two levels: \textit{interpretative} level and \textit{descriptive} level \cite{willig2017interpretation, malterud2016theory, gilgun2015beyond, ngulube2015qualitative, james2013seeking, giorgi1992description}. Researchers play a central role in interpretation, which lies at the core of QDA \cite{ngulube2015qualitative, flick2013sage}. Their key task is perform interpretation, contruct codes that reflect their targeted questions and data. This involves asking questions such as: \textit{What is the concern here? How intense or strong is it? What reasons are given or can be reconstructed? With what intentions or purposes?} Codes generated during coding are thus not neutral; they are interpretive acts shaped by the researcher’s positionality and the broader context \cite{charmaz2006constructing}. 
In another level, the \textit{descriptive} level, researchers identify basic information without interpretation, staying as close as possible to participants’ accounts.


There is potential for humans and LLMs to play complementary roles at different levels of QDA, allowing human sensemaking and interpretation to be preserved through better task delegation \cite{jiang2021supporting}. LLMs excel at processing large-scale data and performing summarization, whereas humans are essential for higher-level interpretation, determining which themes and meanings should emerge during the coding process. To date, however, little research has examined how to design workflows that appropriately support human involvement by considering the analyst’s role in the process.




In summary, we contribute to the advancement of LLM-assisted QDA research by examining theoretical strategies for trustworthiness, particularly transparency and human involvement, and by exploring how to design tools that leverage LLM automation while maintaining methodological rigor. In particular, questions about human involvement, including when humans should intervene and how they prefer to be involved, remain open. Addressing these questions by examining real users’ practices is essential for designing trustworthy LLM-powered QDA workflows.

\section{Design Requirements}
To identify DRs for a trustworthy LLM-powered workflow for QDA, we first conducted a narrative literature review, from which we derived initial DRs. Unlike systematic reviews, narrative reviews provide a semi-systematic and flexible approach to synthesizing knowledge across studies \cite{baumeister1997writing, Sukhera2022NarrativeReviews, snyder2019literature}. We then conducted formative interviews with eight QDA practitioners to refine and finalize all DRs.

\subsection{Narrative Literature Review}
\subsubsection{Method}
We examined 10 papers spanning three areas: trustworthy AI systems \cite{díazrodríguez2023connectingdotstrustworthyartificial, krishna2025facilitating}, trustworthiness theory in QDA \cite{ahmed2024pillars, amin2020establishing, lincoln1985naturalistic, enworo2023application}, and recent work on transparent and conscientious LLM use in QDA \cite{nyaaba2025optimizing, schroeder2025largelanguagemodelsqualitative, wiebe2025qualitative} and prompting techniques \cite{wu2022aichainstransparentcontrollable}. The results have been summarized in Table \ref{tab:initial-design-insights} in the Appendix. We report the key insights below.

\subsubsection{Results}

\paragraph{\textbf{DR1: Automating the whole pipeline transparently through prompt chaining}}
To improve efficiency, automation should extend across the entire analysis pipeline. We adopt thematic analysis \cite{braun2006using, maguire_doing_2017} as the reference pipeline, a widely used QDA method that progresses from text to codes and from codes to themes. Automation can be realized through prompt chaining \cite{wu2022aichainstransparentcontrollable}, a variation of Chain-of-thought (CoT) prompting where the coding task is decomposed into subtasks (e.g., text-to-codes, codes-to-themes) that LLMs complete sequentially, with each output feeding into the next stage. Because QDA is lengthy and complex, structured input and outputs are essential for transferring intermediate results, reducing manual data management \cite{wiebe2025qualitative}, and ensuring traceability across stages \cite{wiebe2025qualitative,díazrodríguez2023connectingdotstrustworthyartificial}. At the same time, automation must minimize error propagation, particularly given the risk of LLM hallucinations in long reasoning chains \cite{díazrodríguez2023connectingdotstrustworthyartificial, chen2025towards}. At the same time, transparency can be achieved through prompt chaining. A transparent reasoning chain makes intermediate steps visible and inspectable \cite{nyaaba2025optimizing}, establishes a basis for users to intervene, interpret, and contribute to critical decisions \cite{ahmed2024pillars,enworo2023application,amin2020establishing,lincoln1985naturalistic}, and ensures that these decisions are documented and revisable for reuse and verification through an audit trail \cite{wolf2003exploring,wiebe2025qualitative,díazrodríguez2023connectingdotstrustworthyartificial}. By embedding such transparent automation, the system can offload routine mechanical tasks while enabling human researchers to concentrate on high-level interpretation \cite{dunivin2024scalablequalitativecodingllms}, thus improving efficiency.


\paragraph{\textbf{Initial DR for human involvement}} 
In addition, ensuring trustworthiness requires sustained human involvement. Users must retain the ability to supervise and evaluate system decisions throughout the process \cite{krishna2025facilitating, díazrodríguez2023connectingdotstrustworthyartificial, wiebe2025qualitative, schroeder2025largelanguagemodelsqualitative, nyaaba2025optimizing}, rather than only at the final stage. Continuous engagement with the data also enables researchers to develop deeper understandings of data \cite{schroeder2025largelanguagemodelsqualitative}.

\subsection{Formative Interview}

\subsubsection{Participants} Our participants were eight QDA practitioners with varying levels of experience. This selection reflects our focus on target users with substantial practical experience, those who treat efficiency as a baseline requirement, and on diverse QDA practices that vary in their rigor.\footnote{We initially tested the system with a broader range of experts and novices, but subsequently narrowed our analysis to experts with substantial practical experience — those who treat efficiency as a baseline requirement.} Participants were recruited through public channels and our professional networks (see Table \ref{tab:formative-participants}). All participants reported prior experience using LLMs in their work. The study was approved by our institutional IRB.

\begin{table}[!htbp]
\centering
\caption{Participants' Demographic Information}
\label{tab:formative-participants}
\vspace{-8pt}
\resizebox{0.48\textwidth}{!}{%
\begin{tabular}{@{}cllllll@{}}
\toprule
\makecell{\textbf{Participant}\\\textbf{ID}}  & \makecell{\textbf{QDA}\\\textbf{Experience}} & \makecell{\textbf{LLM}\\\textbf{Use}} & \textbf{Education} & \textbf{Expertise} & \textbf{Occupation} \\
\midrule
P1  & $>$10 yrs  & Daily & PhD degree & Design                  & Designer, Lecturer\\
P2 & 6--10 yrs  & Monthly & Master's degree & UX Design               & UX Researcher  \\
P3  & 6--10 yrs  & Weekly & Master's degree & Service Design                  & Consultant    \\
P4 & 6 yrs      & Daily  & PhD degree & HCI + ML                & Applied Scientist \\
P5 & 4--6 yrs   & Daily  & PhD degree & Public Policy           & Policy Researcher  \\
P6 & 4 yrs      & Daily & Master's degree  & Gender and Health       & Project Coordinator \\
P7 & 2 yrs      & Daily  & PhD Student & Software Engineering    & NA \\
P8 & \textless1 yr   & Daily & PhD degree & Computer Engineering              & R\&D Manager\\
\bottomrule
\end{tabular}%
}
\end{table}

\subsubsection{Design Probe} To ground the interviews in insights from our literature review, we developed a design probe to guide the interviews. A probe is intended to provoke participants to reflect on and articulate their experiences, feelings, and attitudes in ways that inspire designers \cite{gaver1999design, Sanders02012014}. Our design is grounded in the initial DRs for efficiency, transparency, and human control that we identified in the literature review. We refer to our probe as MindCoder 1.0, a mock-up created in Figma (see Figure \ref{fig:mindcoder-v1}). The coding workflow aligns with the key stages of thematic analysis \cite{maguire_doing_2017}: after uploading data, users proceed through a step-by-step coding process, receive AI-generated code suggestions, validate and interpret these outputs, and then click the “save and generate” button to move to the next stage. The process ultimately generates a codebook in PDF format for download and subsequent use.

\subsubsection{Procedure and Data Analysis} The interviews were conducted via Zoom. After obtaining permission to record the session, the instructor presented the key features of the design probe. Participants were then invited to share their experiences, explain how they would like to interact with the proposed workflow, and suggest improvements to better support their needs in QDA. After the study, we transcribed the recordings. Using a thematic analysis approach \cite{maguire_doing_2017, braun2006using}, two authors independently conducted an initial coding of several interviews, then compared and discussed their interpretations to resolve discrepancies. The resulting themes were further reviewed by the research team. After reaching consensus, one author completed the full coding of the data.

\subsubsection{Results} The interviews enabled us to finalize five more DRs.

\paragraph{\textbf{DR2: Preserving human agency -- humans should not be constrained to a linear, forced process}}
Our design probe described a linear process; however, participants emphasized the need for an iterative process in which they could step in and out at any point and determine when to start or stop. The tool should support user control over aspects such as \textit{“when to use”, “what to use it for”, “which stage and analysis to use it for,”} and \textit{“how much and what type of data to view”} (P3). Participants also expressed the desire not only to edit and wait for the next stage but also to modify the LLM’s groupings according to their own interpretations. For instance, if the LLM did not meet their expectations, users should be able to revise their prompts so that the results change for the entire step and its downstream outputs (P1). As P4 explained, \textit{“Like with ChatGPT, you can point out what went wrong in the previous stage and adjust.”} In addition, the system-provided stages did not reflect all situations. Many participants noted that they do not follow such staged processes; instead, they typically perform only two stages, such as coding and identifying themes (P2, P3). Moreover, the boundaries between layers are often unclear. Participants emphasized that they should be able to write prompts to define the level of specificity in naming themes (P7).

\paragraph{\textbf{DR3: Positioning humans as interpreters, not validators -- system should handle validation tasks.}}
We found that our design probe assigned many mechanical tasks to users, particularly: 1) validating the text against the original data to ensure that no changes were introduced and that important information was not omitted, and 2) manually triggering LLM generation at each step, which required repeatedly waiting for responses. For the first, it was especially concerning for participants who rely on LLMs in their daily work. For example, P1 emphasized the need for constant quality checks due to hallucination issues, noting that LLMs often \textit{“lie to him.”} Similarly, P8 expressed the importance of achieving high code coverage, explaining that he did not want to risk missing any data and having to conduct another round of checks. For the second, P8 highlighted the frustration of waiting for complex prompts to generate, observing that the procedure felt unnecessarily forced. As he explained, \textit{“I find ChatGPT really useful because with just a few simple prompts, I can get a decent output—it gets the job done. Otherwise, I don’t even bother using an LLM.”} These mechanical tasks distracted participants from higher-level interpretation, leading them to view our process as inefficient and troublesome. Clearer task delegation — where LLMs handle mechanical operations and humans focus on interpretive judgments — is therefore needed.


\paragraph{\textbf{DR4: Supporting flexible engagement levels — humans should be able to decide how deeply to engage.}}
Our design probe currently only allowed a fixed level of engagement. Participants emphasized that they should be able to decide the extent of their interpretive involvement depending on the task at hand. If users only wanted to use the tool to generate initial ideas, the system should not force them to engage in in-depth interpretation at every step. In other words, interpretation options should remain available but optional. For example, P2, a UX researcher, explained: \textit{“Just as a starter thought, it [the final PDF] then prompts internal conversation, with other team members adding what they see as the opportunities.”} Building on these early points, she described how she would then manually \textit{“go into the data herself”}. Other scenarios suggested by participants included \textit{efficiently bringing insights back to the team for group discussion} (P1, P2, P3, P4, P5), \textit{identifying a starting point from a high-level overview or supporting early ideation during initial rounds of analysis} (P1, P2, P5), and \textit{teaching junior researchers how to think like a researcher} (P1). Each of these activities demanded varying degrees of interpretive effort. 



\paragraph{\textbf{DR5: Actively encouraging human participation -- system should actively nudge humans to reflect, not assuming it.}}
Since greater control and flexibility are required, it risks replacing the sensemaking process — users may accept results directly without engaging in interpretation. As P4 cautioned, \textit{“Sensemaking is a process that should not be replaced.”} Deep engagement involves reading the data carefully, revising, and rethinking \cite{maguire_doing_2017, braun2006using}. For instance, P1 often engages in reflective practices, such as asking LLMs to self-criticize by adopting a \textit{“critical lens of a highly organized project manager”}. Likewise, the system should encourage users to adopt a reflective stance. This can be supported through nudge-based designs, such as prompts for reflective memo writing, guidance for interpretation, reminders that codes are generated by LLMs rather than humans, and clear signals of tasks that require human input. In addition, the system should communicate its uncertainties and prompt users to reflect, rather than allowing them to assume its outputs are correct by default.






\paragraph{\textbf{DR6: Capturing reusable analytical insights -- system should preserve reflection and interpretation for reuse.}} Our design probe will generate a final PDF that includes the primary codebook to record key decisions made by humans on the basis of LLM results. Participants suggested that this feature could be further enhanced, as many downstream tasks require reusing outcomes and interpretations across different scenarios. For example, this should include the coding structure (P6) and analysis notes (P3). P3 noted that her interpretations often \textit{“become parts of a presentation that [she] would replay back to clients or partners.”} In addition, P4 noted that quickly generating such records can reduce the "repetitive work" of organizing customer feedback analysis data into documents for his company’s weekly meetings. This highlights the importance of establishing an efficient way to preserve intermediate interpretations for further usage.

\section{MindCoder System}

Informed by six DRs, we designed MindCoder, a trustworthy LLM-powered workflow. Below, we introduce MindCoder's usage scenario, key features and implementations.


\subsection{MindCoder Workflow \& Usage Scenario}
\begin{figure*}[!htbp]
  \includegraphics[width=\textwidth]{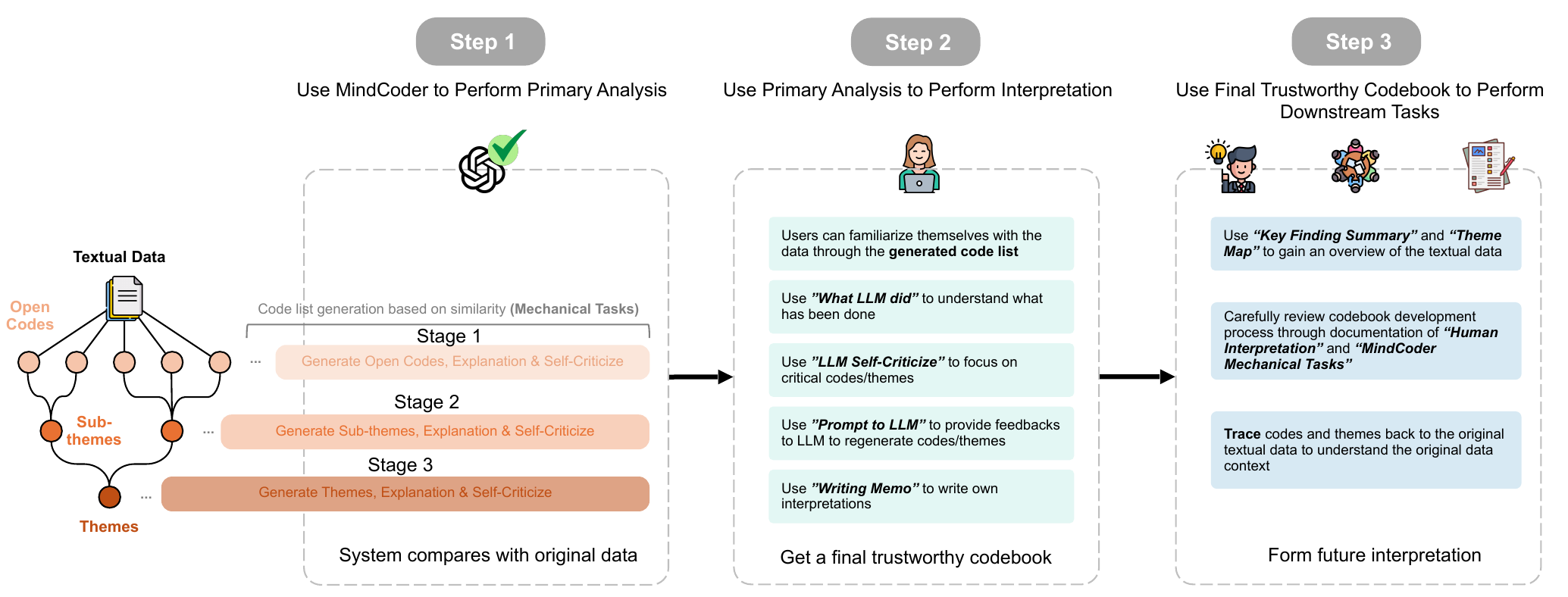}
  \caption{\textbf{MindCoder: A Trustworthy LLM-Powered Workflow for Qualitative Analysis.} Users can perform QDA through the following steps: Step 1. Use MindCoder to conduct primary analysis. Step 2. Build on the primary analysis to perform sensemaking and interpretation, and save the results in a trustworthy codebook. Step 3. Leverage the trustworthy codebook to support downstream tasks, such as group discussions or secondary interpretations.}
  \label{fig:workflow}
\end{figure*}

Suppose Alice, a UX researcher, is gathering customer feedback to share with her team. To assist with coding and analysis, Alice decides to use MindCoder — after carefully removing all customer privacy information. Figure~\ref{fig:workflow} illustrates how MindCoder can be incorporated into her analysis workflow.

\begin{figure*}[!htbp]
    \centering
    \includegraphics[width=\linewidth]{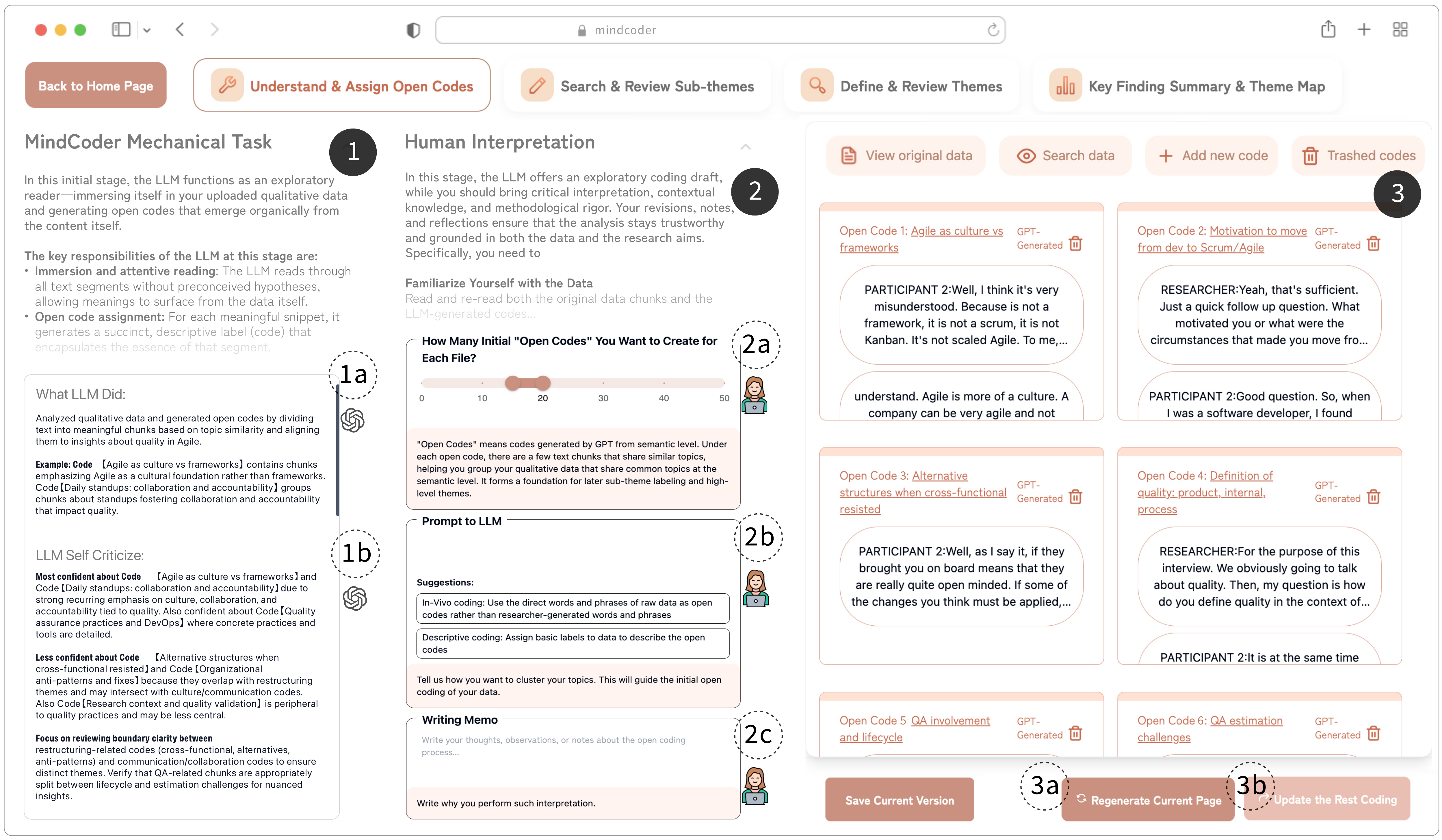}
    \caption{\textbf{Interface for "Stage 1: Generating Open Codes" in "Step 1" of Figure \ref{fig:workflow}}. \textbf{(1) MindCoder’s Mechanical Task}: the LLM (1a) reports “What LLM did” during open coding and (1b) provides self-critique. \textbf{(2) Human Interpretation}: the user (2a) specifies the number of open codes to generate, (2b) writes prompts to regrouping clusters and assign open code names—automatically updating the rationales for new grouping in (1a) and (1b)—and (2c) adds their interpretive memos. \textbf{(3) Displayed Output}: MindCoder presents the LLM-generated open codes. (3a) Users can regenerate codes based on prompts from (2b), and (3b) the LLM can update subthemes and themes to reflect new groupings.}
    \label{fig:codes}
\end{figure*}

\begin{figure*}[!htbp]
    \centering
    \includegraphics[width=\linewidth]{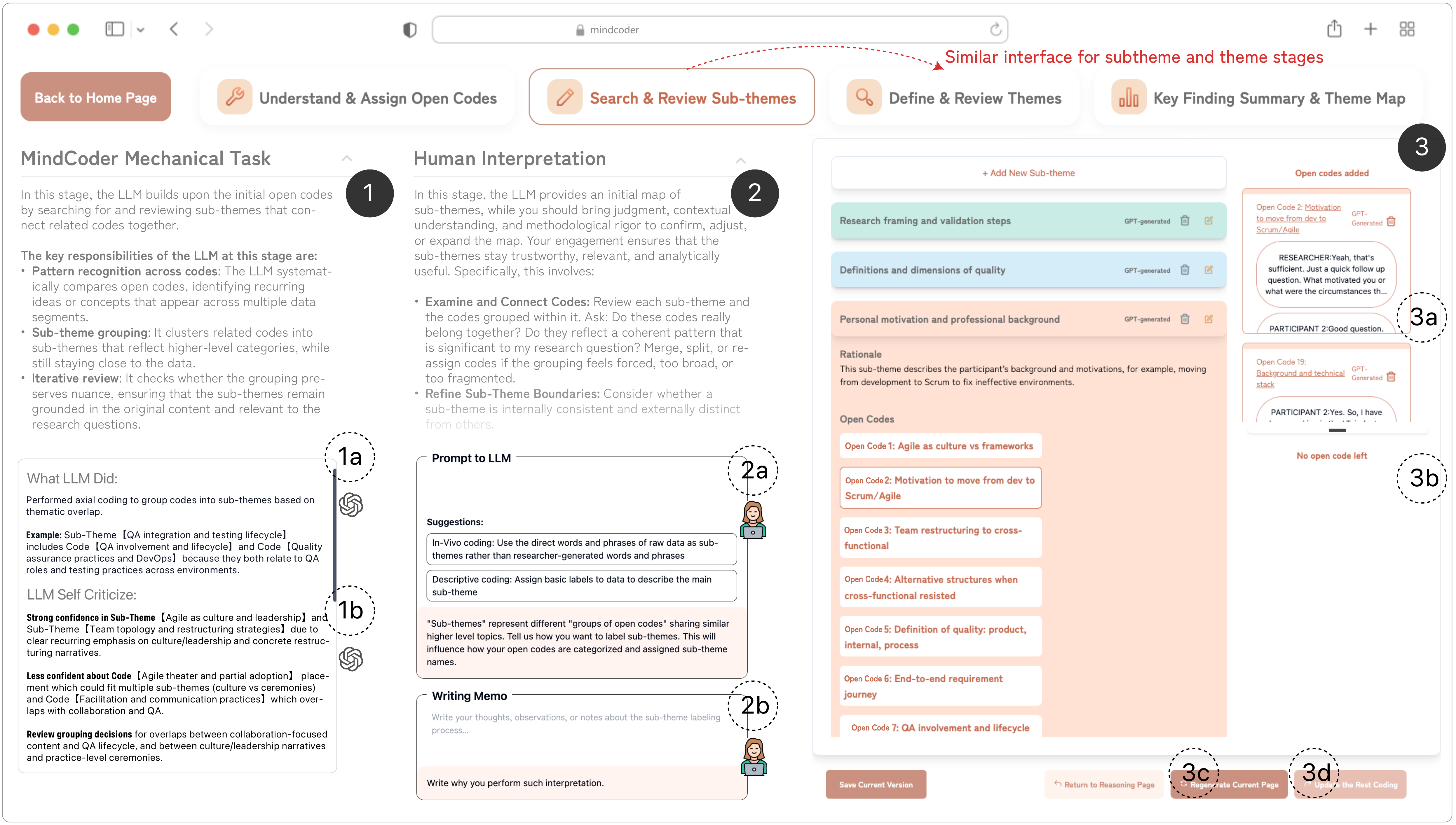}
    \caption{\textbf{Interface for "Stage 2: Generating Sub-themes" in "Step 1" of Figure \ref{fig:workflow}}. (1) \textbf{MindCoder’s Mechanical Task}: the LLM (1a) reports “what it did” during open coding and (1b) provides a self-critique. (2) \textbf{Human Interpretation}: the user (2a) writes prompts to regroup clusters and assign subtheme names—automatically updating the rationales in (1a) and (1b)—and (2b) adds interpretive memos. (3) \textbf{Displayed Subthemes}: MindCoder presents the LLM-generated subthemes. (3a) Users can review open codes under each subtheme, (3b) see any ungrouped codes (none in this example), (3c) regenerate results based on prompts from (2a), and (3d) update the rest of the coding to align with the new subthemes.}
    \label{fig:subtheme}
\end{figure*}

In \textit{Step 1}, Alice uploads the data into \mindcoder and optionally customizes the research questions (see Figure \ref{fig:final-rqs}). She then proceeds directly to coding. Because her dataset is relatively large (over 10,000 words), the system requires approximately 15 minutes to generate a preliminary analysis. During this time, Alice takes a short coffee break. When she returns, the preliminary results are already displayed in the MindCoder interface: Open Codes (Figure \ref{fig:codes}), subthemes and themes (Figure \ref{fig:subtheme}).

In \textit{Step 2}, Alice uses MindCoder’s preliminary results to guide her interpretation. For instance, when assigning open codes, she first reviews the explanation of MindCoder’s mechanical task. She then examines “What LLM Did” to understand how the system grouped the data, followed by “LLM Self-Criticize” to assess its confidence levels across codes—“most confident,” “less confident,” and “ambiguous.” The last category requires Alice’s focused review. Through this process, she develops an initial understanding of the LLM’s rationales of current coding.

After building this initial understanding, Alice goes to the “Human Interpretation” guideline to clarify her tasks. She can then adjust the number of open codes, prompting the LLM to regroup them automatically. Alternatively, she may use “Prompt to LLM” to craft new instructions—for example, “make the themes shorter” or “generate code names with more original keywords.” These prompts are used to regenerate the codes, and were saved in the “Prompting History” for later usage. 

During the process, Alice can also rename codes, search or add new ones, and track revisions: any modified code is labeled “user edited” rather than “GPT-generated”. She records her interpretations—such as code changes, memos, rationales, or reasons for revision—in the “writing memo.” Once satisfied, she proceeds to the subtheme and theme interfaces to engage in sensemaking and higher-level interpretation following a similar process. Finally, by using “view original data” (Figure \ref{fig:view data} and \ref{fig:view data2}), she can examine highlighted excerpts linked to their sources, compoared with original data to ensure quality by the system, which helps her contextualize and refine her edits. The interface also displays coverage information.

In \textit{Step 3}, Alice can view the key findings summary in the interface, tracing original data displayed. Once satisfied with the results, she clicks “Save this version” to store them as the finalized output. Using this final report, she can apply the completed codebook to trace back to the original data and share the results with her team.

\subsection{Key Features}

\subsubsection{Automated and Transparent Reasoning Chain}

\paragraph{Preliminary Analysis through QDA Reasoning Chain}  

To align with DR1, \mindcoder enables users to generate codes, subthemes, and themes based on similarity-level grouping. The system achieves this by providing a prompt chain — each step is performed by the LLM, and the output is passed to the next step as the input.  

\paragraph{Validating Results against Original Data before Presentation to Users}  

To align with DR1 and DR3, the validation task is performed by the system. \mindcoder achieves this primarily by using the data from the first step. At all stages, the system provides links so that users can check the original data. In addition, the system calculates the coverage of the coded text and maps these segments back to the original text, ensuring that users can easily see which parts have been coded and which have been missed.  

\paragraph{Flexibly Selecting Stages for Human Intervention}  
To align with DR2, users should be able to intervene at any stage, as the output of each stage is displayed on the interface after data generation. For additional agency, users can exit the workflow at any time by saving the final results as a PDF.

\begin{figure*}[!htbp]
    \centering
    \includegraphics[width=0.98\linewidth]{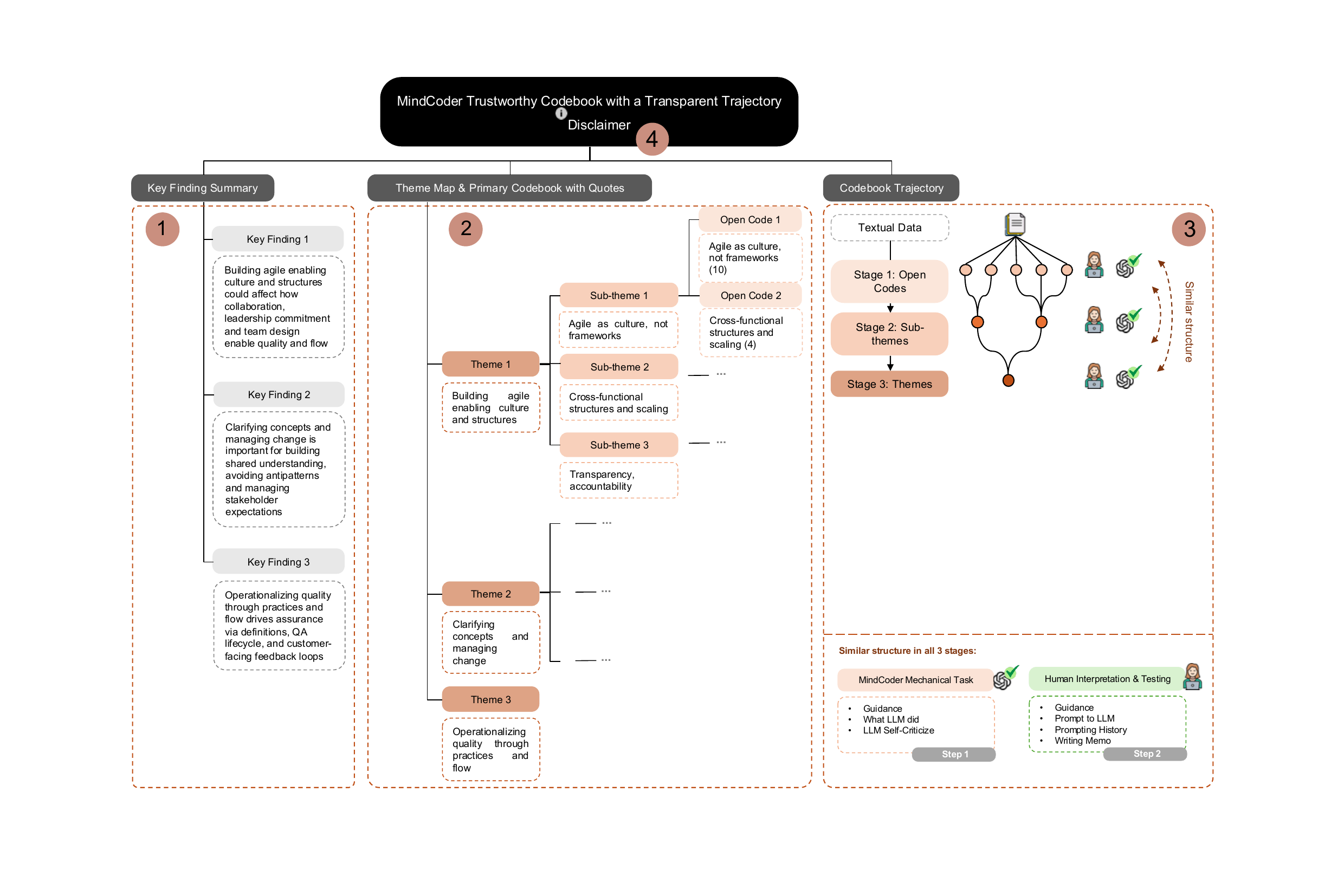}
    \caption{The Structure of MindCoder’s Trustworthy Codebook with a Transparent Trajectory. \textbf{1) Key Finding Summary} shows high-level synthesized takeaways for readers; \textbf{2) Theme Map \& Primary Codebook} shows structured codes, sub-themes, and themes with exemplar quotes \textbf{3) Codebook Trajectory} shows stepwise progression from open codes to themes, showing user's interpretation and reflection in each step; \textbf{4) Disclaimer} clarifies the limitations of AI-generated content and reinforces interpretive role of human analysts. We show a sample trustworthy codebook in supplementary materials.}
    \label{fig:codebook}
\end{figure*}

\begin{figure*}[!htbp]
    \centering
    \includegraphics[width=\linewidth]{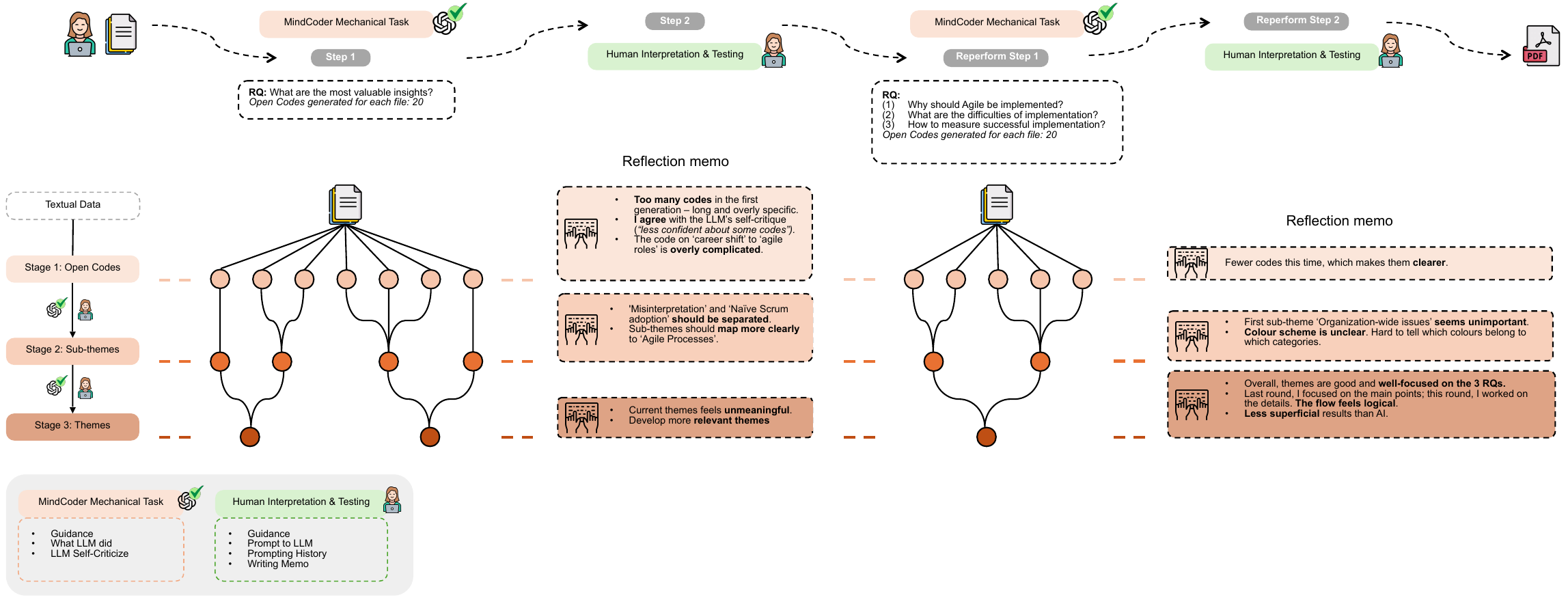}
    \caption{An example of user's reflection trajectory (P10 in final study). This figure shows how this user worked with MindCoder in two rounds: starting with dense, detailed codes, then refining them into clearer and more focused themes through new questions and chosen settings.}
    \label{fig:user-trajectory}
\end{figure*}

\subsubsection{Data Exploration}  

\paragraph{Viewing and Editing Codes, Subthemes, and Themes}  
In each interface, users can interactively view the codes, subthemes, and themes. To align with DR2, \mindcoder enables users to edit each stage by adding, deleting, or renaming codes, subthemes, and themes. Users may also add additional text content to existing codes when they identify more relevant material in the raw data. All such edits trigger updates in subsequent steps of the reasoning chain.

\subsubsection{Reflection and Interpretation}  
To align with DR3, DR5, \mindcoder provide a list of nudges and reflection tools to support human to perform reflection and interpretations. 

\paragraph{Nudges -- Explaining Mechanical Tasks to Support Human Sensemaking}  
To fulfill DR5, \mindcoder is designed to reduce the burden of human sensemaking and to encourage active intervention by making the LLM’s underlying reasoning process explicit. It explains \textit{what it did} and the rationale for why specific texts were chosen and grouped as codes, subthemes, and themes. In addition, it provides self-criticism of its confident, less confident, and ambiguous codes, thereby guiding users’ interpretive focus.  

\paragraph{Clarifying Human Responsibility}  
To align with DR3 and DR5, \mindcoder clarifies the human’s responsibility for interpretation, emphasizing active engagement rather than passive review, and specifies the tasks that users need to perform.

\paragraph{Prompt to LLM}
To align with DR2, users can refine the reasoning chain by adjusting parameters or by writing direct prompts at each stage. During open coding, \mindcoder allows users to specify the number of open codes and to provide targeted prompts (e.g., “use original terms in the text”) that guide the LLM to generate codes and themes aligned with their preferences. 

\paragraph{Prompting History}  
To align with DR6, all tested prompts are saved in the final PDF to support future downstream use and replication.

\paragraph{Writing Memo}  
To align with DR3 and DR6, users should record their interpretations throughout the process. They may revise prompts, adjust codes, or generate new insights that prompt them to write reflections. All such reflections can be documented as memos, which are saved in the final PDF for future use.

To align with DR4, all of these processes are optional, and users can decide whether to skip or perform them.

\subsubsection{Final Trustworthy Codebook with Trajectory}

To align with DR6, \mindcoder generates a final trustworthy codebook that includes not only the primary codebook—containing codes, themes, their corresponding quotes, the visualized theme map, and a summary of key findings—but also documentation of each role’s responsibilities and tasks at each stage as part of the codebook development trajectory. Figure \ref{fig:codebook} describes the codebook structure, while Figure \ref{fig:user-trajectory} describes a real user trajectory from our user study.

\paragraph{Disclaimer}  
To align with DR5, the final PDF generated by the LLM includes a disclaimer noting that the codes are partially generated by GPT. This also serves as a nudge to remind both analyst users and downstream users to maintain full responsibility for reviewing, modifying, or rejecting the system’s outputs and to engage with the content critically.   

\paragraph{Theme map}  
To align with DR6, to provide further usage, users can view the final meaning hierarchy through a theme map and decide which visualizations to retain. This visual abstraction enables users to grasp thematic relationships at a glance and to detect potential overlaps or inconsistencies before exporting the results as a final PDF.

\paragraph{Codebook Development Trajectory}  
To align with DR6, this documentation includes prompting history, the number of open codes selected, and reflection memos, all of which can be exported to a well-formatted PDF within one second. Such traceable documentation enhances transparency, enabling analyst users to present their interpretations to the team and allowing downstream users to leverage the codes and judge whether to trust them based on how humans reflected on the codebook.

\subsection{Prompt Design}

\label{sec:prompt_design}

The prompt design mainly consists of two parts:
\textit{1) Reasoning Chain for Preliminary Analysis}: For automation purpose in DR1, we constructed a complete reasoning chain to generate the preliminary analysis. The key steps of the reasoning chain that perform basic thematic grouping based on text similarity. This is achieved by applying prompt chaining to decompose the complex reasoning process into smaller subtasks \cite{wu2022aichainstransparentcontrollable, promptingguide2023promptchaining}, a variation of Chain-of-Thought (CoT) prompting \cite{wei2022chain}. The main steps of the reasoning chain follow the logic of thematic analysis, progressing from raw data to codes, subthemes, and final themes \cite{maguire_doing_2017, saldana2021coding}, thereby supporting automation of the entire pipeline.
\textit{2) Explanation Prompts}: For reflection purpose in DR3 and DR5, prompts that explain \mindcoder’s mechanical tasks to support human sensemaking in each stage, including generating “What the LLM Did” and “LLM Self-Critique.” 

Moreover, we applied best practices of prompt design\footnote{\url{https://platform.openai.com/docs/guides/prompt-engineering}}\footnote{\url{https://www.promptingguide.ai/introduction/tips}} to structure the prompts. For each prompt, we specified the task description (step-by-step prompting), parameters (e.g., number of codes, uploaded data), requirements for code structure and format, output format, and few-shot examples in JSON to ensure accurate input and output. Below, we introduce the main task descriptions to provide a high-level understanding of what each prompt does. For more detailed prompt designs, please refer to Appendix~\ref{sec:prompt_list}. 

\subsubsection{Generating Open Codes.}
In this stage, the original qualitative data is grouped into open codes based solely on text similarity. Users can determine the level of specificity by adjusting the number of topics, and they may further guide generation by setting label conventions — for example, requesting more specific or more interpretive codes. The prompting process proceeds in three steps. First, the text is divided into chunks (i.e., \textit{ Analyze the raw text content from the uploaded data and segment it into meaningful chunks based on topic similarity. Each chunk must retain the exact original text without any modifications.}) Next, these chunks are grouped based on shared and similar topics (i.e., \textit{Organize these chunks into multiple codes (as specified by the ‘Number of codes’), ensuring that each code groups only text chunks with similar topics.}) Finally, a basic name is assigned to each group (i.e., \textit{Assign a short name to each code that reflects its basic meaning.})

\subsubsection{Generating Sub-themes and themes.}
In this stage, \mindcoder groups similar open codes based on their similarity and then assigns subtheme names. First, it performs grouping of codes in last stage (i.e., \textit{Group similar codes based on high-level thematic overlap.}) Each group is then assigned a primary name (i.e., \textit{For each group, assign a primary name that best captures the main subtheme represented by the grouped codes.}). To ensure that the results remain at a basic level with minimal interpretation, we added the following prompt (i.e., \textit{Ensure subtheme names are descriptive and specific, incorporating key concepts, terms, and entities drawn directly from the content, without interpretation.}) The generated subthemes are sent to the LLM to generate higher-level themes. At this stage, the process is also aligned with the research questions by asking the LLM to do so (i.e., \textit{When possible, assign research questions to guide the generation of these high-level theme names}).

\subsubsection{Generating Key Finding Summary \& Visualization.} To distill the key findings from the insights, we asked the LLM to summarize the relationships between each theme and the research questions, if provided.
\textit{1. Examine the uploaded codebook and source data to extract and summarize key findings aligned with each theme, focusing on how they address the research questions. 2. Incorporate original themes, subthemes, or codes to support each finding.} To generate visualizations, \mindcoder sends the meaning hierarchy to the LLM, asking it to generate mindmap visualization as a “DOT language graph designer.”

\subsubsection{Generating Sensemaking Nudges}
In each stage, we provided prompts to support human sensemaking, particularly through “What the LLM Did” and “LLM Self-Critique.” These prompts summarize the main actions (e.g., “generate open codes by dividing text into meaningful chunks and grouping them by similarity”). We also included examples to clarify codes and asked the LLM to identify its most confident groups, codes, and themes. In addition, for each theme, the description summarizes and explains the main content of its associated themes and codes.

\subsection{Implementation}
We implemented \mindcoder as a web application, including a React-based frontend with a Node.js backend. With frontend server handles the user data input and interactions, request response via backend through OpenAI's GPT-5 model API \footnote{\url{https://platform.openai.com/docs/models/gpt-5}}. 
We selected GPT-5 because it supports a 400,000-token context window and up to 128,000 output tokens, ensuring that the model can handle relatively large text data through a single API call. The reasoning mode was set to minimal\footnote{\url{https://platform.openai.com/docs/guides/latest-model}}, as this setting provided sufficient performance while reducing response time and maintaining high data coverage.

At each step, the system combines user-uploaded data (e.g., original text, user-written prompts, parameters) with the predefined prompt template and the coding data from that step (e.g., codes, subthemes, themes) before sending them to the GPT-5 model. All data are represented as JSON objects (e.g., themes, subthemes, codes, and GPT responses) for easy data management and accurate generation.

For most prompts in the reasoning chain, we applied standard text-based prompting, using natural language as both input and output. For visualization tasks, we used the DOT language to represent the coding hierarchy. DOT\footnote{\url{https://graphviz.org/doc/info/lang.html}} is a graph description language used to define nodes, edges, and subgraphs. These elements were explicitly specified within the prompt, and the LLMs were instructed to generate DOT representations of the hierarchical coding structure. Further details about the visualization prompt design are provided in Appendix.

\subsection{Technical Evaluation}
We performed a technical evaluation to ensure that \mindcoder outputs accurate content and can be trusted. Reliable task performance is essential so that users can trust the correctness of GPT-generated results. A key challenge introduced by \mindcoder’s whole pipeline automation is the risk of hallucination and error propagation. This arises because \mindcoder’s prompt spans four stages, each of which processes large text inputs and outputs, in addition to long prompt instructions (approximately 1000 words per stage, plus uploaded text that may extend to several thousand words). Accurately handling such large inputs can be challenging, even for GPT-5. In particular, we employed multiple strategies to ensure the accuracy of the prompts (see Appendix~\ref{sec:prompt_list} for details) and iteratively refined them until strong performance was achieved. We report the evaluation methods and results below.

\subsubsection{Metric} 
To evaluate whether the content remains stable and unchanged from text to codes (i.e., coverage rate)\footnote{We evaluated performance only for the transformation from text to codes, not for later stages, because once the first stage is ensured, subsequent stages replace their codes within these text groups before displaying them to users and therefore avoid introducing LLM generation errors.}, we apply the Jaccard similarity\footnote{\url{https://www.ultipa.com/docs/graph-analytics-algorithms/jaccard-similarity}}. 
Formally, let $W_i$ denote the set of words at stage $i$ and $W_{i+1}$ denote the set of words at stage $i+1$. 
The Jaccard similarity between these two stages is defined as:

\begin{equation}
J(W_i, W_{i+1}) = \frac{|W_i \cap W_{i+1}|}{|W_i \cup W_{i+1}|}
\end{equation}

This metric measures the proportion of overlapping words relative to the original data in the first stage (from text to codes), thereby capturing the extent to which the content remains consistent from stage~$i$ to stage~$i+1$.

\subsubsection{Setup}
To cover a broad range of data types, we collected six types of common qualitative data: blog data \cite{schler2006effects}, email data \cite{Cohen2015Enron}, meeting transcripts \cite{Janin_ICSI_2004}, Quora answers\footnote{\url{https://www.quora.com/}}, Wikipedia\footnote{\url{https://www.wikipedia.org/}}, and Stack Overflow posts\footnote{\url{https://stackoverflow.com/questions}}. For each data type, we manually collected 12 documents of 2,000–4,000 words, roughly equivalent to a 20-minute TED talk at an average speaking rate of 140 words per minute\footnote{\url{https://tfcs.baruch.cuny.edu/speaking-rate/}}. This length is sufficient to capture diverse meanings, contexts, and themes, while remaining manageable for the LLM. After collecting the text data, we manually cleaned them by removing irrelevant content, unusual characters, and formatting issues. The dataset can be found in supplementary materials. For each dataset, we run reasoning chain using GPT-5 for five times to reduce randomness and report the average results below. The dataset is attached in supplementary material.

\subsubsection{Results}
Our evaluation results show that the system achieves high similarity across datasets. The average Jaccard similarity is 89.6\%, with 91.4\% for blogs, 96.1\% for emails, 66.3\% for meeting transcripts, 96.2\% for Quora, 93.8\% for Wikipedia, and 93.6\% for Stack Overflow. Among these, conversational data such as meeting transcripts exhibit the lowest similarity. This finding is reasonable, as transcripts often contain verbal text and unstructured expressions that should not be carried forward in the coding process. By contrast, relatively structured data (e.g., Wikipedia, blogs, emails, Quora, and Stack Overflow) demonstrate stronger retention and stability. 

In summary, our reasoning chain is reliable in the transformation from text to codes, and in subsequent stages the system reuses this data, thereby avoiding error propagation throughout the longer reasoning process. This design not only ensures efficiency — by preventing redundant system or human checks at each stage — but also enhances trustworthiness, as users can rely on the stable LLM outputs throughout the pipeline to support higher-level interpretations.


\section{User Evaluation}

To evaluate \mindcoder, we conducted a user study with 12 participants serving as analysts and an expert validation with 2 external expert evaluators serving as stakeholders. Specifically, our evaluation questions are:

\begin{itemize}
\item \textbf{RQ1:} How does the \textit{trustworthy LLM-powered workflow} MindCoder satisfy diverse user needs and influence acceptance among analyst users?
\item \textbf{RQ2:} To what extent can the \textit{trustworthy LLM-powered workflow} MindCoder enhance downstream users’ perceptions of the trustworthiness and quality of QDA results?
\end{itemize}

\subsection{Study Design}

\subsubsection{Participants}
We recruited 12 participants (see Table~\ref{tab:final-study-participants}) through university mailing lists and public Slack channels. All participants had at least a basic level of understanding or practice in QDA, along with some experience using LLMs. Among them, six self-reported as intermediate, four as experts, and two as novices. Each participant received compensation equivalent to $15.60$ USD for completing the study, in accordance with our local reimbursement rate for user studies. This study has been approved by our institutional IRB.

\subsubsection{Conditions}
For comparison, we selected Atlas.ti with its AI coding function, a state-of-the-art commercial CAQDAS powered by OpenAI’s model, as the most reasonable benchmark. We did not choose ChatGPT or other conversational LLM tools, as they are not designed for QDA and differ substantially in functionality from \mindcoder.

\begin{itemize}
    \item \textbf{Baseline: Atlas.ti AI Coding} — a commercial, state-of-the-art LLM-integrated software that incorporates OpenAI’s model. It operates one-click code generation, assigning open codes alongside the original text and allowing users to perform post-hoc validation against the source material. A demonstration video is available on its website\footnote{\url{https://atlasti.com/ai-coding-powered-by-openai}}. 
    \item \textbf{MindCoder:} The complete MindCoder workflow and system. 
\end{itemize}

Latin Square Design. We employed a within-subject design in which each participant used two tools on two different datasets. To control for order effects, we applied a Latin square design so that no participant completed the tasks or used the tools in the same sequence (see Table~\ref{tab:final-study-participants}).

\subsubsection{Tasks and Datasets} For each task, participants used one of the tools to develop a codebook within approximately 25 minutes, following three steps: (1) briefly familiarize themselves with the data, (2) use the system AI function to generate primary coding and (3) review, reflect on, and interpret the system-generated codes. At the end of the task, all codes and reflections were saved as a final report. To encourage engagement and ensure the quality of the final results, participants were instructed to treat the exercise as a real analysis scenario and to approach it carefully and reflectively. They were also informed that their codebooks would contribute to future qualitative research.

For each task, codebook development was based on one of the datasets we provided. These were curated from two publicly available interview datasets: one on software quality \cite{alami2022scrum} and one on qualitative data teaching \cite{curty2024teaching}. Each dataset consisted of two interviews of approximately 1,000 words\footnote{This length was chosen to simulate real-world scenarios in which multiple data sources are analyzed. In addition, given GPT-5’s long generation time, this size allowed participants to complete the task within 1.5 to 2 hours.}. We selected interview data because it reflects common conversational formats such as focus groups and meetings. To balance complexity and ensure participants were not hindered by content difficulty, we simplified portions of the text into plainer language.

\begin{table}[!t]
\centering
\caption{Participants' Demographic Information and Study Arrangement}
\label{tab:final-study-participants}
\renewcommand{\arraystretch}{1.3}
\resizebox{0.5\textwidth}{!}{%
\begin{tabular}{@{}cllllll@{}}
\toprule
\textbf{PID} &  \makecell{\textbf{QDA Exp.}} & \textbf{LLM Use} & \textbf{Background} & \textbf{Education} & \textbf{Session 1} & \textbf{Session 2}\\
\midrule
P1  & Intermediate  & Daily   & Data Science \& AI   & Undergraduate Student & Task 1 $\times$ MindCoder & Task 2 $\times$ Atlas.ti AI Coding \\
P2 & Intermediate  & Daily & Psychology \& Management  & PhD Student  & Task 1 $\times$ MindCoder & Task 2 $\times$ Atlas.ti AI Coding\\
P3  & Expert  & Monthly  & Psychology                  & Master's Degree  & Task 1 $\times$ MindCoder & Task 2 $\times$ Atlas.ti AI Coding \\
P4 & Expert   & Daily   & HCI           & PhD Student  & Task 1 $\times$ Atlas.ti AI Coding & Task 2 $\times$ MindCoder \\
P5  & Expert     & Weekly  & Psychology     & Undergraduate Degree  & Task 1 $\times$ Atlas.ti AI Coding & Task 2 $\times$ MindCoder \\
P6 & Intermediate      & Daily   & Social science               & Undergraduate Student & Task 1 $\times$ Atlas.ti AI Coding & Task 2 $\times$ MindCoder\\
P7 & Intermediate      & Daily   & Computing AI       & Master Student & Task 2 $\times$ MindCoder & Task 1 $\times$ Atlas.ti AI Coding\\
P8 & Novice      & Daily   & Data Science   & Master Student & Task 2 $\times$ MindCoder & Task 1 $\times$ Atlas.ti AI Coding\\
P9 & Novice      & Daily   & Information systems                     & Undergraduate Degree    & Task 2 $\times$ MindCoder & Task 1 $\times$ Atlas.ti AI Coding\\
P10 & Expert      & Daily   & Genetics \& Social Science    & PhD Student   & Task 2 $\times$ Atlas.ti AI Coding & Task 1 $\times$ MindCoder\\
P11 & Intermediate     & Weekly   & Audit Associate    & Bachelor's degree  & Task 2 $\times$ Atlas.ti AI Coding & Task 1 $\times$ MindCoder \\
P12 &  Intermediate     & Daily   & Business Administration    & Undergraduate Student  & Task 2 $\times$ Atlas.ti AI Coding & Task 1 $\times$ MindCoder \\
\bottomrule
\end{tabular}%
}
\end{table}

\subsubsection{Study Procedure}
The 1.5-hour online user study was conducted via Zoom. At the beginning, the instructor introduced the study, explained its purpose, and asked participants to complete a pre-study questionnaire. With their consent, the sessions were recorded, and participants were asked to share their screens to facilitate interaction. Before each task session, participants received training on the two software tools (approximately 10 minutes) and a short period (approximately 3 minutes) to review and familiarize themselves with the dataset. The instructor was also available to answer any questions as needed. They then performed the coding tasks using each tool in their assigned sequence. Throughout the process, participants were encouraged to reflect, take notes, and think aloud. 

\begin{figure}[!t]
    \centering
    \includegraphics[width=1.01\linewidth]{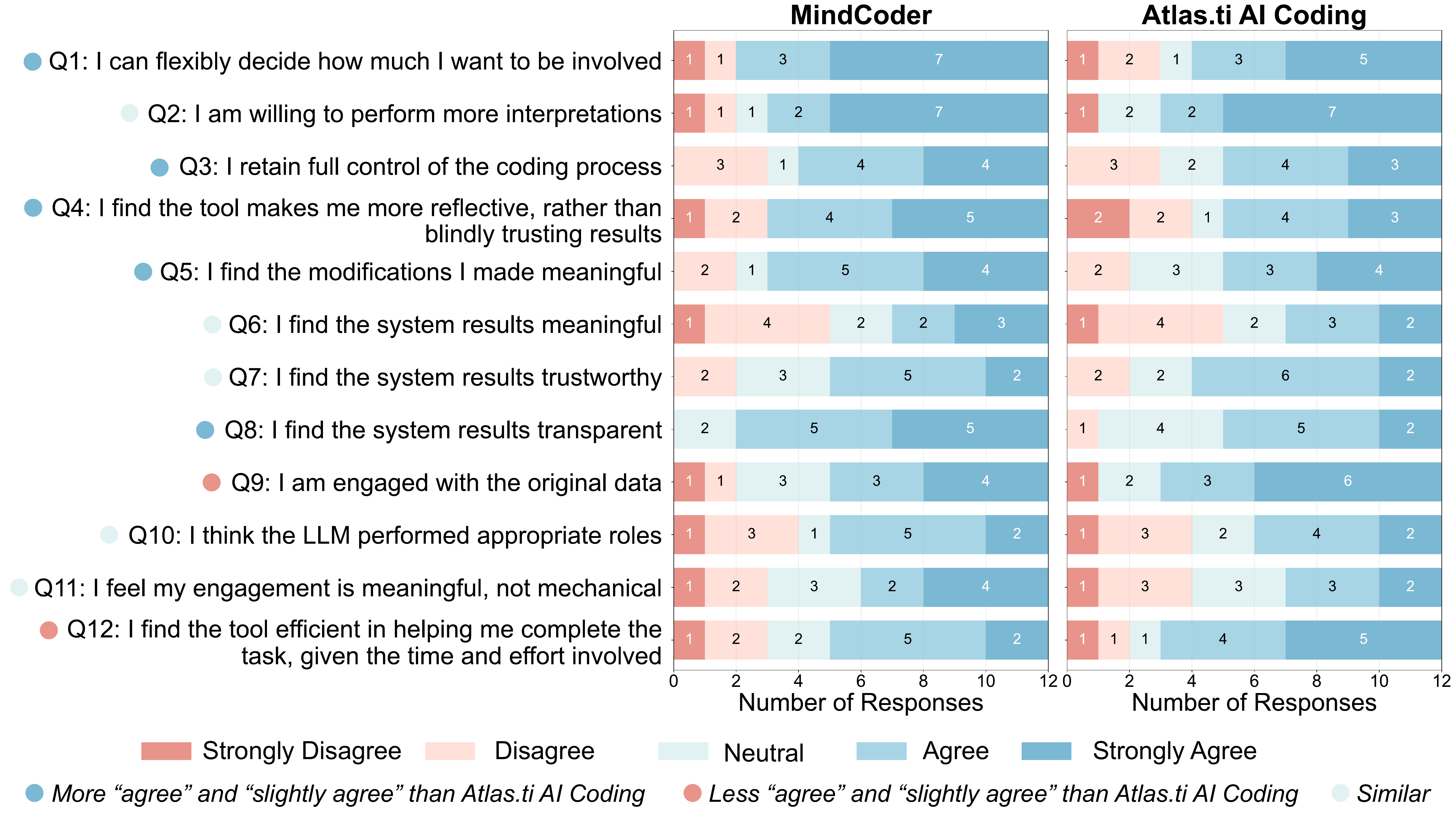}
    \caption{Comparison of Likert-Scale Responses in Post-study Questionnaire}
    \label{fig:comparison}
\end{figure}

At the end of the study, participants completed a post-study questionnaire in which they rated their subjective experiences with each platform on a 5-point Likert scale. They evaluated 12 statements (e.g., “I find…”, “I think…”) addressing usage experience such as flexibility, controllability, and meaningfulness (see Figure~\ref{fig:comparison})\footnote{We excluded NASA-TLX and SUS measures to avoid bias in the comparison, as Atlas.ti is a commercial software with full automation, and participants may already be more familiar with this type of CAQDAS. As a result, it would predictably achieve higher usability scores and lower cognitive load than our prototype.}. In addition, the instructor conducted a semi-structured interview to gather qualitative feedback on their experiences. To ensure objective evaluation, participants were not informed which tool was designed and developed by us, and they stopped screen sharing while completing the post-study questionnaire.

\subsubsection{Data Analysis} To analyze user feedback from the post-study interviews and think-aloud sessions, we transcribed the audio recordings and conducted a thematic analysis \cite{maguire_doing_2017, braun2006using}. The first author (who conducted the study) read and reviewed the data, performed open coding, and identified initial themes. A second author, an expert in qualitative research, reviewed the coding and candidate themes and contributed to interpretation through a series of discussions aimed at resolving discrepancies and achieving consensus. Finally, the analysis was presented to the entire research team for member checking and shared interpretation.

\subsection{Results}

\subsubsection{Questionnaire Results} Although the t-tests did not reveal statistically significant differences between the two tools, likely due to the small sample size, we report the descriptive statistics below to highlight potential trends.

\paragraph{When is MindCoder slightly better?} More participants perceived MindCoder slightly better than Atlas.ti mainly in terms of Flexibility (Q1), Reflexivity (Q4), the meaningfulness of modifications (Q5) and the Transparency (Q8). For Q1, 10 participants agreed that MindCoder allowed them to flexibly decide how much they wanted to be involved, compared with 8 for Atlas.ti. For Q4, 9 participants reported that MindCoder encouraged them to be more reflective about AI-generated results, whereas only 7 expressed the same view for Atlas.ti. For Q5, 9 participants felt their modifications in MindCoder were more meaningful compared to 7 in Atlas.ti. For Q8, 10 participants indicated that MindCoder enhanced transparency, while 7 reported the same for Atlas.ti. 

\paragraph{When two tools are similar} A majority of participants (9/12) indicated that they would be willing to use both tools for additional interpretation (Q2), and most reported feeling that they retained full control (Q3: 7–8 of 12). However, only 5 participants found the system-generated results meaningful in both systems prior to their interpretation (Q6), whereas more participants considered their own modifications meaningful in MindCoder than in Atlas.ti (9 vs. 7; Q5). Interestingly, participants were not particularly sensitive to trustworthiness: 7–8 of the 12 rated both systems as trustworthy (Q7).

\paragraph{When is Atlas.ti AI Coding slightly better?} While both tools were generally rated as efficient, Atlas.ti’s AI Coding received more agreement ratings on efficiency (9 vs. 7 for MindCoder). This is reasonable given that Atlas.ti provides rapid code generation, whereas \mindcoder, powered by latest GPT-5, has a longer generation time and performs additional steps, such as deriving subthemes and themes, before producing results. In addition, more participants reported stronger engagement with the original data when using Atlas.ti (Q9: 9 vs. 7), which can be explained by Atlas.ti’s design as a CAQDAS platform that affords users direct exposure to the raw text.

\subsubsection{Key Findings (KFs)} We report key findings related to analysts’ usage experiences and their acceptance below.

\paragraph{\textbf{KF1: \mindcoder fosters deeper interpretive engagement and critical reflection.}}
Overall, while most participants did not consider system-generated results meaningful for QDA on their own, they reported that MindCoder enabled them to engage more deeply with the results, make modifications with the system’s suggestions, producing outcomes they perceived as more meaningful.

Participants attributed this to the descriptive and detailed nature of MindCoder’s codes (P6, P7), which they found more explainable compared to Atlas.ti’s shorter topic labels (P4). The descriptive content also sparked ideas they had not previously considered (P10). As one participant explained, the two systems felt suited to different tasks: MindCoder supported interpretation, while Atlas.ti supported labeling (P2, P7). P2 noted, \textit{“I did feel like I was trying to do slightly different tasks with the MindCoder versus Atlas.ti. With the MindCoder, I took more of a summarizing perspective, while with Atlas.ti, I focused more on making sure each code corresponded to the correct sentence.”}

The descriptive nature of MindCoder’s codes helped participants stay more engaged and reflective. For instance, P10 emphasized that while it could not \textit{“replace human thinking”}, it can serve as an \textit{“alongside AI partner”} that provided additional perspectives, or a second coder to confirm or challenge his thoughts: \textit{“Coder 1 is me, Coder 2 is the AI. I can do it myself first, and then we can compare separately using the logic. In such a way, MindCoder is like a second person.”} Similarly, P7 contrasted the two tools, suggesting that Atlas.ti risked reducing critical engagement: \textit{“With MindCoder, I look at the whole content and make sure all the key information is summarized. With Atlas.ti, I think it makes me a lot lazier, because all of the codes are already there. I just need to look at them, so I don’t engage as much in critical thinking… I had to put in more thought and engage more with the MindCoder.”}

At the same time, participants raised concerns about whether highly descriptive codes always supported sensemaking. P8, a novice, questioned whether they had the expertise to evaluate such detailed outputs: \textit{“Because MindCoder is very in-depth and detailed, to the extent that I question whether I really need to intervene too much. But for Atlas.ti, because it’s so simple, I had a bit more mental capacity to think about the themes myself.”} This suggests that preferences may vary by expertise: while some participants appreciated descriptive codes for prompting reflection, novices sometimes preferred simpler codes to reduce cognitive load.

\paragraph{\textbf{KF2: MindCoder provided flexible control for varied analytical habits and tasks}}
While both tools were perceived as offering substantial control, some participants especially valued \mindcoder’s flexible prompting at intermediate steps. At a basic level, participants noted that \mindcoder provided more explanations and allowed them to edit prompts. As P4 said, \textit{“With MindCoder, I get more explanations, and it also allows me to edit the prompt. But with Atlas.ti, since it’s a commercial product, I can’t see or edit the prompts. In Atlas I didn’t have much control, I just used toggles. Overall, I think MindCoder gives me more sense of control, while Atlas.ti feels more like a black box.”}

We also observed differences in analysts’ coding behaviors, particularly in how novices and experts exercised control. Junior analysts tended to begin with high-level interpretations (P7, P8, P11). For instance, P11 explained, \textit{“In business analysis, you usually start with a high-level problem statement and then break it down into subcategories. I actually prefer starting from the highest level.”} Similarly, P7 remarked, \textit{“I think I would like it to start from themes, because the codes were initially very overwhelming for me. If I had gone through themes first, I could have understood it better.”} By contrast, experts (P3, P4, P6) preferred starting from the codes, either to pursue a data-driven approach (P4) or to follow more formal analytic routines: \textit{“I feel like if you start from themes, you already have some preconceived assumptions… you might try to force-fit the raw data rather than the other way around.”} (P6)


Beyond differences in coding strategies, participants emphasized that flexible control also helped them accommodate a variety of analytical tasks. For example, P7 and P9 described using \mindcoder to support reading research papers, while P4 appreciated its ability to provide high-level directions that “relieved [her] reading burden”. Similarly, P2 envisioned scenarios where analysts might prioritize broad themes over fine-grained details, \textit{"One example I’m thinking of is a diary study, where every day we ask people to reflect on what happened, like what happened at work or their interactions with colleagues. In that kind of study, we’re not focused on the minute details, just the general concepts. So for that type of study, I think the \mindcoder would be more useful."} In contrast, participants perceived Atlas.ti as more suitable for straightforward topic labeling, given its non-interpretative nature (P5), or for qualitative, detail-oriented analysis such as grounded theory, where its ability to extract quotes within seconds was valued (P2). 

This high-level control also created opportunities for deductive coding. While our initial expectation was that \mindcoder would primarily support inductive analysis, we observed that P10 employed it deductively by starting with predefined themes and theories. After reading the data, he formulated three research questions and then asked the LLM to generate high-level themes and group the data according to those questions. As he explained: \textit{“The overall themes are very focused on my three questions. Maybe as a user, the first view should be naive (like ‘what are the key points’) and only in later iterations ask more specific questions. I think it’s pretty cool. I don’t know much about Agile and Scrum, but it does a better job than if I were to do it myself—less superficial with the AI.”}

\paragraph{\textbf{KF3: MindCoder’s perceived trustworthiness is implicit and varies across users}}

Participants did not describe trustworthiness as a direct evaluation criterion. Instead, they inferred it indirectly through qualities such as the usefulness of information (P7), transparency of rationale (P1), or traceability to original data (P2, P8). For instance, P7 remarked that \mindcoder was more trustworthy than Atlas.ti because it delivered richer insights: \textit{“I trust \mindcoder more because depending on the time and the need to get good information out of it, I would go for MindCoder over Atlas.ti, because the one-word information [in Atlas.ti] was not sufficient for me.”} Similarly, P2 valued transparency through traceability: \textit{“Both are trustworthy. I think it’s transparent enough that I can see the codes correspond to what is said in the original transcripts.”} For P1, the system’s rationale provided reassurance, even if not always consulted in detail: \textit{“Maybe when I’m doing the analysis, I won’t really read everything about the rationale, maybe because of time constraints, but if I need to understand more about the AI’s analysis, then it’s very useful for me.”}

In contrast, P4 emphasized that trustworthiness was less of a concern for her, since she preferred to maintain control through manual analysis: \textit{“I still need to produce my own version, so trust doesn’t matter as much because I’ll manually process the transcript anyway.”}

\paragraph{\textbf{KF4: MindCoder’s reports were appreciated for clarity and practicality}}
We found that \mindcoder’s visualizations and PDF reports that included the analysis process and the primary codebook were highly appreciated by participants (P1, P8, P9). P1 highlighted the clarity of the visualization: \textit{“It gave me a chart that was very visually straightforward, understandable, and readable.”} P9 similarly valued the reporting function as a practical foundation: \textit{“I can envision saving time when generating reports. For example, after finishing my qualitative work, like interviewing friends, I could feed the transcripts into MindCoder and get a solid output to work with as a base.”}

\paragraph{\textbf{User Acceptance and Challenges}}

We found that QDA users differed in their acceptance of the system. The workflow tended to be more appealing to individuals with expertise in both QDA and LLMs, as their technical and methodological knowledge shaped their priorities differently. For example, participants P1, P2, P4, P6, P7, P10, P11, and P12, who were proficient in both domains, valued \mindcoder’s potential. They described the system as insightful, meaningful, flexibly and useful.

In contrast, P3, P5, P8, and P9 expressed concerns and frustrations. These were largely linked to familiarity with traditional CAQDAS systems (P3), the steep learning curve of \mindcoder (P5, P8, P9), a preference for closer engagement with the original data (P3), perceptions of neutrality in traditional tools (P3), reduced sense of closeness to data when mediated by AI (P9), and delays due to long API regeneration times (P9). These factors were viewed as advantages of traditional CAQDAS.


\subsection{External Expert Evaluation}
To evaluate the outcomes, 12 pairs of codebooks (24 in total) produced by \mindcoder and Atlas.ti for 12 participants, we invited two external experts. 
Each expert spent approximately four hours on the evaluation. We chose external experts rather than conducting the evaluations ourselves in order to ensure objectivity and consistency in the ratings.

\subsubsection{External Experts Recruitment}
The two expert evaluators were recruited from different disciplinary backgrounds: psychology (Evaluator 1, E1) and data science (Evaluator 2, E2). E1, a PhD student with four years of QDA experience, regularly applies QDA in academic research and represents a methodological perspective grounded in qualitative traditions. In contrast, E2, an undergraduate in data science with about two years of QDA experience, brings a technical background and a more flexible approach to QDA in research contexts. This combination enabled us to capture evaluations from both methodological and technical standpoints.

\subsubsection{Evaluation Setup}
Before reviewing the codebooks, the experts were asked to familiarize themselves with the datasets. We then introduced them to the codebooks generated by 12 users and asked them to evaluate the results on a 10-point Likert scale (10 is better) across six questions (Q1-Q6) representing four dimensions of trustworthiness (credibility, confirmability, dependability, transferability). The questions included aspects such as credibility, reflection of the original data, clarity of intermediate steps, and overall understandability. Each question corresponded to one or more trustworthiness dimensions—for example, Q1 and Q2 mapped onto credibility, while Q3 mapped onto confirmability. The specific rating questions and full mappings are presented in Figure \ref{fig:final-evaluation}.

To avoid bias, the evaluators were not asked to assess the codebooks in the same session, thereby preventing their judgments from influencing one another. Neither evaluator had prior exposure to the project or the software tools. Moreover, they were blinded to the source of each report: they were not told which tool was developed by us and which is commercial one. 

\subsubsection{Results}

\begin{figure}[!htbp]
    \centering
    \includegraphics[width=\linewidth]{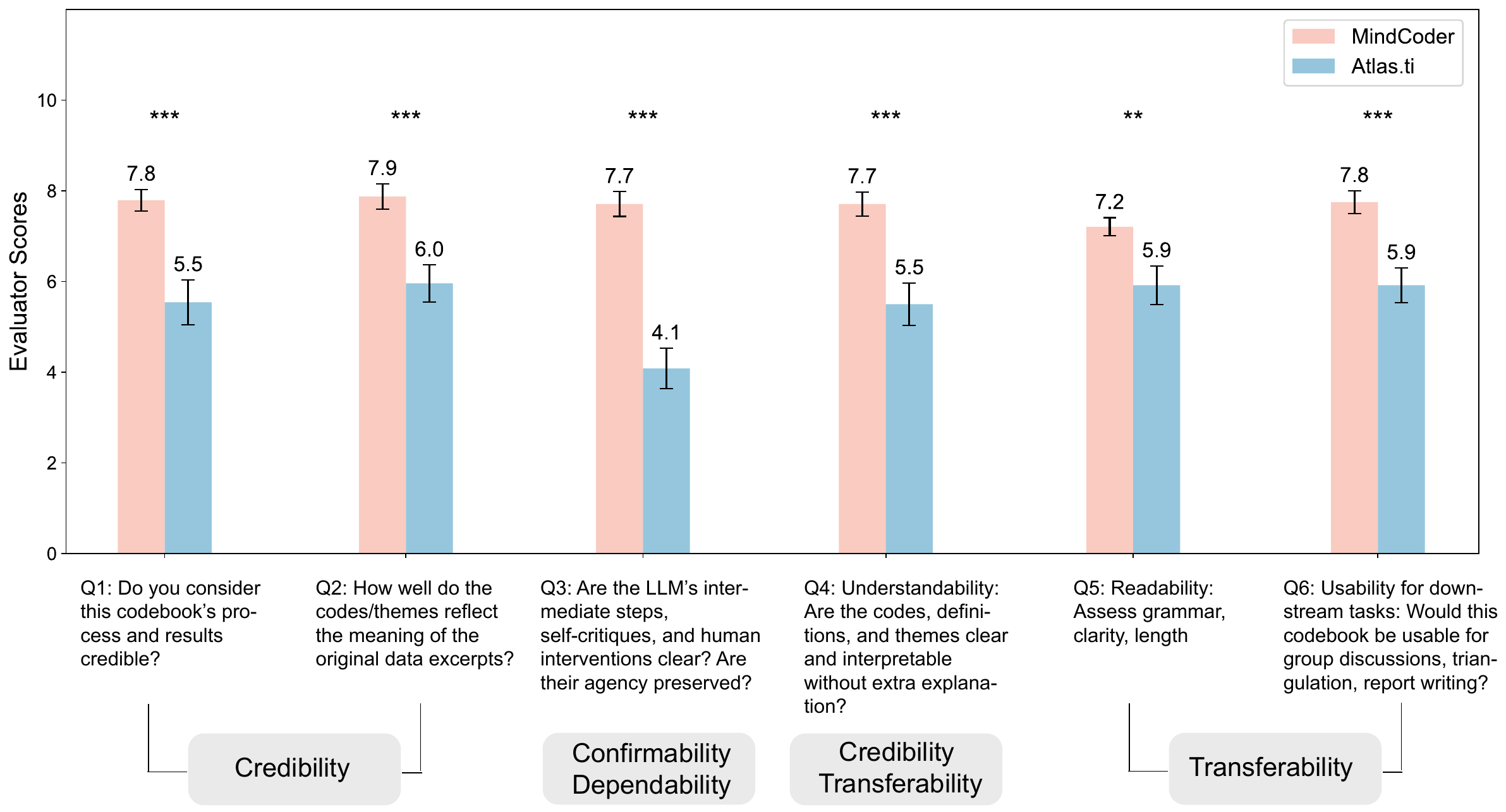}
    \caption{Expert Evalution Scores on 24 codebooks (12 from MindCoder vs. 12 from Atlas.ti) generated from user evaluation. *** means $p< .001$, ** means $p<.01$.}
    \label{fig:final-evaluation}
\end{figure}

Overall, our preliminary evaluation indicated that experts rated \mindcoder's outcomes significantly higher than Atlas.ti across all six questions (paired t-test, all $p < .01$), suggesting higher perceived trustworthiness among participants (Figure \ref{fig:final-evaluation}). We next report the preliminary qualitative feedback from experts.

\paragraph{Credibility}
Both experts found that MindCoder’s outputs were credible, data-grounded. E1 noted that the tool “captures concepts and relationships in detail,” while E2 emphasized its “strong code-to-theme alignment.” Together, these features produced structured codes, subthemes, and themes that reflected the dataset’s core ideas. E1 further observed that “the codes generated are detailed, structured, and clear.” Similarly, E2 described MindCoder’s results as systematic and thorough, though “occasionally redundant or over-conceptualized,” and suggested that they could “benefit from sharper distinctions between overlapping codes.” Moreover, E1 suggested that the context can be kept so that the original data's flow is not broken, \textit{"We shouldn’t override or interrupt it [the converation dynamic in data]. Since I mostly do psychology research, I think if we try to record the transcript, we can’t just extract parts of it, we can’t only pull out certain sections."} In contrast, Atlas.ti was viewed as overly concise, often reduced to label, like codes lacking supporting evidence and broader logic. As E2 mentioned, \textit{“While Atlas.ti captures broader meaning, it sometimes reduces precision.” }

\paragraph{Transferability}
Both experts found that MindCoder can produce richly detailed and well-structured outputs that are good for downstream tasks. E2 thinks that the codebook "had a clear structure, concise definitions, and a visual theme map" and "the roles of LLM and analyst made the codebook usable for downstream tasks". 
However, its outputs were sometimes considered too repetitive or difficult to navigate in visual maps by E2. E1 emphasized that a balance and tradeoff between level of details and concise is needed. As she suggested, \textit{"I think sometimes it’s a bit redundant for content to be listed in such detail. It’s a trade-off or balance, you want to preserve the details, but sometimes capturing the essence with short topics or titles makes it easier to access later, especially when researchers go beyond this project and look back."} By contrast, Atlas.ti was valued for being concise and efficient but was also critiqued for being too brief or fragmented, which limited its usefulness for cross-sample synthesis.  Furthermore, E1 criticized Atlas.ti’s “poor labeling and color use,” which undermined the reliability of reuse and interpretation. E2 also observed that “some merged codes appear broad and risk losing nuance from the original data.” Overall, Atlas.ti’s outputs were seen as fitting better for early brainstorming stages but brings the risk of confusion when the original data was unavailable.

\paragraph{Confirmability} We expected that \mindcoder's outcome can somehow provide other people ability to perform group member checking, and understand how analyst thinks, so that can decrease the biases. This could be reflected by memo record of user reflection. Experts, especially E2, noted that MindCoder makes human–system relationships clearer and more transparent. For example, he mentioned: \textit{“LLM–human interaction is evident through self-critiques and memos”}. 
By contrast, he thinks that Atlas.ti's Codebook: \textit{“Analyst–LLM interaction is minimal, leaving agency less explicit.”} Thus, although the LLM–analyst relationship is somewhat visible, intermediate reasoning steps and human interventions remain only partially transparent.

\paragraph{Dependability}
Both experts found that \mindcoder's output can provide somewhat reproducibility, since the codes area clear, while "a bit misordered, redundant, or not mutually exclusive." (E1)
While sometimes the stucture of themes and subthemes' relationship seems not correct, but "since it provides step-by-step record, it will be good to do the steps with mistakes". (E1)

\subsection{Summary of Answers to RQs}
\subsubsection{RQ1: How does MindCoder satisfy diverse user needs and influence acceptance among analyst participants?}

To address RQ1, \mindcoder primarily supports analyst users by enabling more meaningful involvement, such as interpretation and critical reflection (KF1 \& questionnaire results). It also offers flexible controls that adapt to diverse user needs and analytical tasks (KF2 \& questionnaire results). Notably, while both systems (MindCoder \& Atlas.ti) were perceived as trustworthy, analysts do not evaluate the system directly in terms of trustworthiness; instead, they rely on indirect measures such as the usefulness of information and the transparency of rationales (KF3 \& questionnaire results). Moreover, \mindcoder's outcomes were considered clearer and more practical (KF4). In addition, \mindcoder was particularly effective for individuals with expertise in both QDA and LLMs, as it provided deeper insights and fostered more meaningful engagement. By contrast, users with less technology background found traditional CAQDAS-integrated AI functions more familiar and easier to learn.


\subsubsection{RQ2: To what extent can MindCoder enhance downstream users’ perceptions of the trustworthiness and quality of QDA results?}

To address RQ2, we found that \mindcoder outperformed the baseline (Atlas.ti AI Coding) across all evaluated dimensions, including credibility, faithfulness to the original meaning, and inclusion of intermediate steps. Its detailed documentation of the coding process, systematic structure, and more precise definitions contributed substantially to credibility, dependability, confirmability, and, to some extent, transferability. 



\section{Discussion}

\subsection{Designing "meaningful" human involvement for LLM-powered QDA}
In this paper, we began with theoretical strategies for establishing methodological rigor in trustworthy LLM-powered QDA, focusing on transparency and human involvement. In addition, we also conducted interviews to gain expert insights regarding how to design human involvement meaningful in applications, from which we summarized into six practical, human-accepted design requirements or DRs. We found that our DR-informed, LLM-powered QDA system supports more practical use by facilitating active interpretation and reflection, accommodating diverse analytical needs and tasks, and producing significantly more trustworthy primary codebooks. 

Overall, building systems that are both efficient and trustworthy takes more than following abstract guidelines, it is important to apply a human-centered way that gathers user feedback about what are acceptable and meaningful. For example, DR3, positioning humans as interpretators, not validators. While references and page numbers provide useful traceability \cite{nyaaba2025optimizing}, it is important to examine how much mechanical tasks like traceability and validation should be allocated to the system versus to human analysts, from a task delegation perspective \cite{jiang2021serendipity}. Because there is no clear boundary between interpretive and mechanical tasks, fully excluding LLMs from interpretive roles may be neither feasible nor desirable. Our study suggests a different possibility for avoid displaced human sensemaking: instead of mirroring the analyst currently using the system \cite{oksanen2025llmcode}, the LLM might mirror another researcher with a different positionality \cite{holmes2020researcher} and perspective, thereby acting as a “second coder” to confirm or challenge users. Our KF1 suggests that appropriate LLM support accepted by users could include (1) human-modified descriptive-level interpretation and (2) reflective nudges that encourage users to reflect, thereby fostering greater human engagement while maintaining automation. Similarly, other DRs should incorporate user feedback to ground the design of human involvement in both theory and practice, supporting more meaningful real-world engagement.


\subsection{Two meanings of "trustworthiness" in LLM-powered QDA}

In our work on LLM-assisted QDA, we identify two distinct forms of trustworthiness.

\subsubsection{Establishing trustworthiness for AI systems in QDA.}
This refers to how analysts perceive the reliability of LLM-generated outputs. We found that most analysts intuitively judge trustworthiness by assessing the meaningfulness and quality of results rather than by examining rationales or reasoning steps. 
Only analysts familiar with LLM limitations valued detailed rationales, and even they preferred direct access to data, referring to rationales only when necessary. For most analysts, then, trustworthiness is a byproduct of useful, high-quality outputs. If the system is doubted, results are often disregarded or underused, limiting opportunities for reflection. This suggests that developers must embed mechanisms of transparency and evidence even if users do not actively consult them, ensuring that analysts always know the evidence is available.

\subsubsection{Establishing trustworthiness for QDA downstream users.}
 This concerns the credibility of the results produced with the aid of LLM systems. For downstream users, such results must be supported with “thick description” and clear evidence of analytic processes — expectations aligned with qualitative research standards of trustworthiness \cite{ahmed2024pillars, lincoln1985naturalistic, amin2020establishing}. Here, trust is explicit: results must be justified and verifiable to be accepted as meaningful.

Taken together, these two perspectives reveal that system-level trustworthiness functions less as an end in itself and more as a tool. By supporting meaningful analyst engagement — even if analysts rarely scrutinize the system’s trustworthiness directly — transparent and careful system design enables analysts to build results that downstream users find trustworthy. Thus, in LLM-assisted QDA, system trustworthiness is best understood as an enabling condition that underpins the explicit trustworthiness required for analyst-generated results.

\subsection{Epistemology and Users}

We developed our design rationale around the needs of QDA practitioners who aim to conduct analyses at a basic to medium level of methodological rigor while maintaining high efficiency. Typical users include UX researchers who analyze user reports on a weekly basis, HCI researchers who employ QDA primarily to evaluate their systems, and managers who rely on analysis results to inform decision-making.

Many of these practitioners adopt a positivist orientation, emphasizing truth-seeking rather than interpretive exploration \cite{blatter2017truth}. For such users, MindCoder supports eliciting truth in the data by automating descriptive tasks such as clustering, open coding, and fact-checking. Analysts can then build interpretation on top of this descriptive foundation. In this sense, MindCoder functions as a scaffold: it enhances trustworthiness at the descriptive layer while deliberately leaving interpretive sense-making to human analysts. While our initial design principle (DR3) aimed for MindCoder to offload mechanical tasks (e.g., grouping and clustering), our first key finding (KF1) revealed that the system actually encouraged users to be more engaged in interpretation. At first glance this might appear contradictory, but it in fact illustrates the intended design effect: by delegating descriptive tasks to the LLM and exposing them transparently, MindCoder created more space for human interpretation.  In other words, efficiency gains at the mechanical level translated into richer involvement at the interpretive level, showing that automation and interpretive depth are not in conflict but can be mutually reinforcing when roles are carefully divided.

This approach is particularly useful in situations such as iteratively refining early analyses by adding more data after the first few interviews, sharing preliminary insights with a team to spark group interpretation, initiating informal discussions, developing storylines, applying existing theories to data, or training novice researchers.

However, interpretivists \cite{lin1998bridging} may find MindCoder more constraining, as its predefined groupings can inadvertently limit interpretive freedom. For these users, MindCoder is better positioned as a second coder, applied after their initial coding to provide additional perspectives without restricting their own interpretive process.

\section{Limitation and Future Work}

This study has several limitations. First, although trustworthiness can be established through many strategies, our design focused only on two: transparency (via showing intermediate outcomes, thick description and audit trail) and human involvement (through reflection and interpretation). While we introduced intermediate steps to enable user interaction, additional strategies, such as triangulation with multiple data sources, remain unexplored. Future research should examine these strategies to strengthen trustworthiness and further deepen human involvement.

Second, the translation of high-level concepts of trustworthiness into design requirements was based mainly on feedback from eight practitioners with a positivist orientation. Other epistemological perspectives, such as interpretivism, may yield different design requirements. However, our results show that the current DR-informed design already satisfies many users’ needs. Future work should therefore engage a more diverse range of participants to capture varied perspectives.

Third, our workflow relied primarily on the GPT-5 model, which showed the best performance for our tasks. However, its long generation times affected the user study and created barriers for participants with less technical familiarity. Although repeated use may reduce this learning curve, since many prompts are reusable, the system is model-agnostic. Future models that handle larger documents more efficiently could significantly improve system performance.

Fourth, we compared MindCoder only with Atlas.ti AI Coding. Because both tools are powered by OpenAI models, the comparison was influenced by factors such as usability and user familiarity. Broader comparisons with other AI-supported software (e.g., Otter.ai, Zoom’s AI Companion) are needed to further contextualize our findings. Moreover, the user study was small in scale, leaving open questions about how larger and more diverse groups would respond to the tool. In addition, our expert evaluation was also preliminary. Future research could therefore include larger studies and develop carefully designed metrics to assess perceptions of trustworthiness more rigorously.

\section{Conclusion}

LLMs are increasingly applied in QDA, yet unvalidated automation and reduced human sensemaking pose critical challenges to meaningful and trustworthy results. This paper addresses trustworthiness through transparency and human involvement. Based on formative interviews, we identified six design requirements (DRs), including ensuring human agency, reserving interpretation for humans, supporting flexible involvement, providing reflective nudges, and enabling outcome reusability. Guided by these DRs, we developed MindCoder, a web-based workflow with a transparent reasoning chain that lets users intervene at intermediate steps while automatically logging analysis trajectories and outputs. A within-subject evaluation with 12 users showed that MindCoder fostered more meaningful engagement and flexible control, and produces more trustworthy analytic outcomes.



\bibliographystyle{ACM-Reference-Format}
\bibliography{reference}

\appendix

\section{Initial Design Requirements}

\begin{table*}[!htbp]
\centering
\small
\caption{Mapping of goals, concepts, design insights, and sources to initial design requirements (DRs).}
\label{tab:initial-design-insights}
\begin{tabularx}{\textwidth}{>{\centering\arraybackslash}m{1.8cm} 
                            >{\centering\arraybackslash}m{1.8cm} 
                            >{\raggedright\arraybackslash}X 
                            >{\raggedright\arraybackslash}p{3.2cm} 
                            >{\centering\arraybackslash}m{1.8cm}}
\toprule
\textbf{Goals} & \textbf{Concepts} & \textbf{Initial high-level design insights} & \textbf{Paper Source} & \textbf{DRs} \\
\midrule
\multirow{8}{*}{Efficiency} 
& Automation & Prompt chaining decomposes a complex task into smaller subtasks. At each step, the LLM is prompted with a subtask, and its response is then used as input for the next prompt. & Wu et al., 2022 \cite{wu2022aichainstransparentcontrollable} & 
\multirow{8}{*}{DR1} \\
& \makecell{Avoid Error\\Propagation} & Verify expected operation and performance enhances reliability and reproducibility of AI-based systems & Diaz-Rodriguez et al., 2023 \cite{díazrodríguez2023connectingdotstrustworthyartificial} & \\
\midrule
\multirow{14}{*}{Transparency} 
& Awareness & Stepwise prompting strategy enhances transparency by systematically generating codes, themes and interpretations & Nyaaba et al., 2025 \cite{nyaaba2025optimizing} & 
\multirow{12}{*}{\makecell{DR1}} \\
& Traceability & Users must be informed of AI systems' capabilities and limitations and are always aware when interacting with AI & Diaz-Rodriguez et al., 2023 \cite{díazrodríguez2023connectingdotstrustworthyartificial} & \\
& Traceability & Keep an audit trail of research decisions, changes, and data analysis processes & Ahmed et al., 2024 \cite{ahmed2024pillars}; Amin et al., 2020 \cite{amin2020establishing}; Lincoln et al., 1985 \cite{lincoln1985naturalistic}; Enworo et al., 2023 \cite{enworo2023application} & \\
& Explainability & LLM interactions should output justifications for key decisions and intermediate steps & Wiebe et al., 2025 \cite{wiebe2025qualitative}; Diaz-Rodriguez et al., 2023 \cite{díazrodríguez2023connectingdotstrustworthyartificial} & \\
\midrule
\multirow{8}{*}{\makecell{Human\\Involvement}} 
& \makecell{Human Agency\\~\& Oversight} & Users must retain the ability to supervise, evaluate AI system decisions with intermediate outputs & Diaz-Rodriguez et al., 2023 \cite{díazrodríguez2023connectingdotstrustworthyartificial}; Ronanki et al., 2025 \cite{krishna2025facilitating}; Wiebe et al., 2025 \cite{wiebe2025qualitative}; Ronanki et al. 2025 \cite{krishna2025facilitating} & 
\multirow{8}{*}{Initial DR} \\
& \makecell{Human-Data\\Engagement} & LLMs foster closeness with data, offering alternative perspectives, challenging existing theories and deepening researcher' understanding & Schroeder et al., 2025 \cite{schroeder2025largelanguagemodelsqualitative} & \\
\bottomrule
\end{tabularx}
\end{table*}

\section{Design Probe}
\begin{figure*}[!htbp]
    \centering
    \begin{subfigure}[t]{0.33\linewidth}
        \centering
        \includegraphics[width=\linewidth]{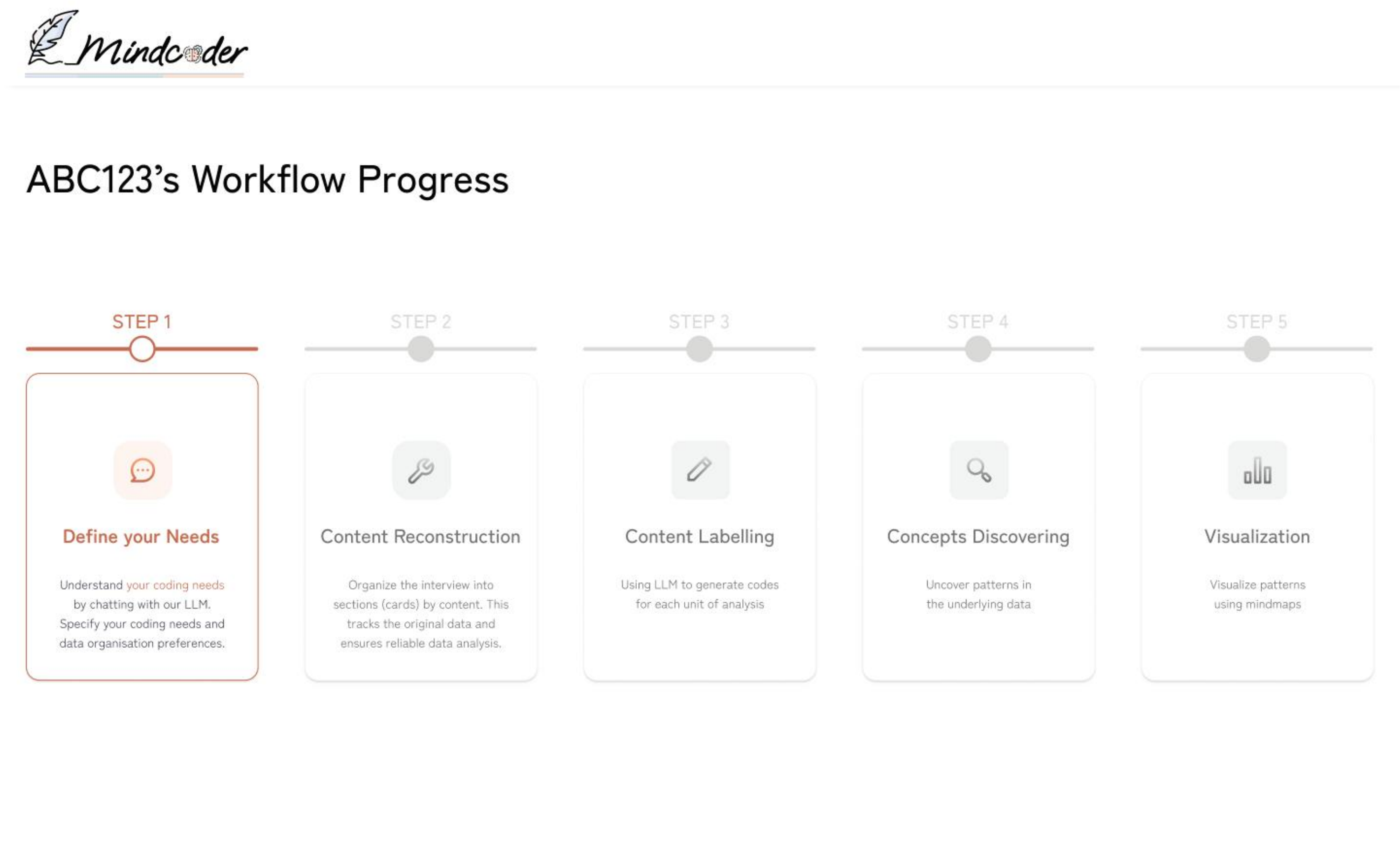}
        \caption{Steps}
        \label{fig:steps}
    \end{subfigure}
    \begin{subfigure}[t]{0.33\linewidth}
        \centering
        \includegraphics[width=\linewidth]{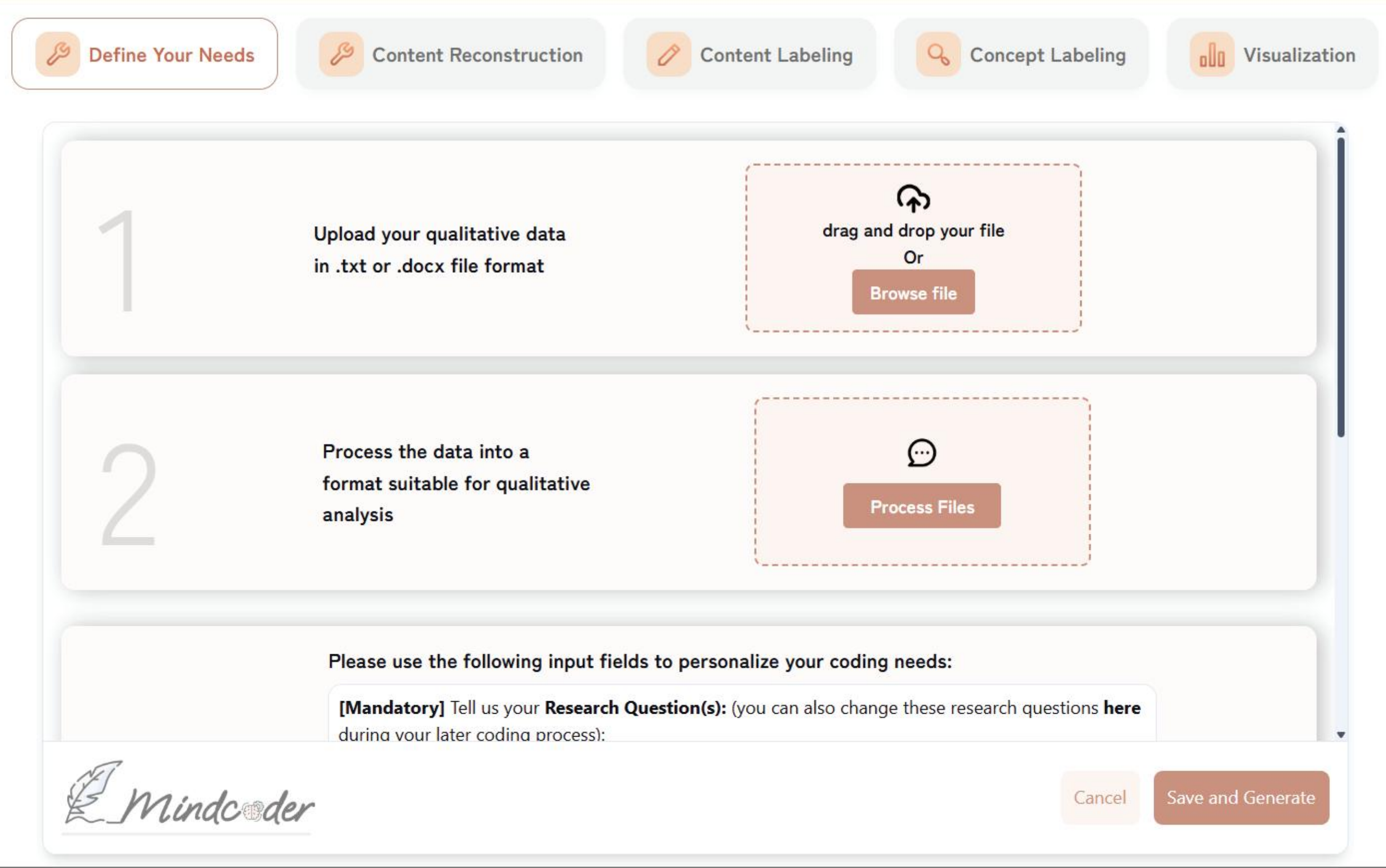}
        \caption{Needs}
        \label{fig:needs}
    \end{subfigure}
    \begin{subfigure}[t]{0.33\linewidth}
        \centering
        \includegraphics[width=\linewidth]{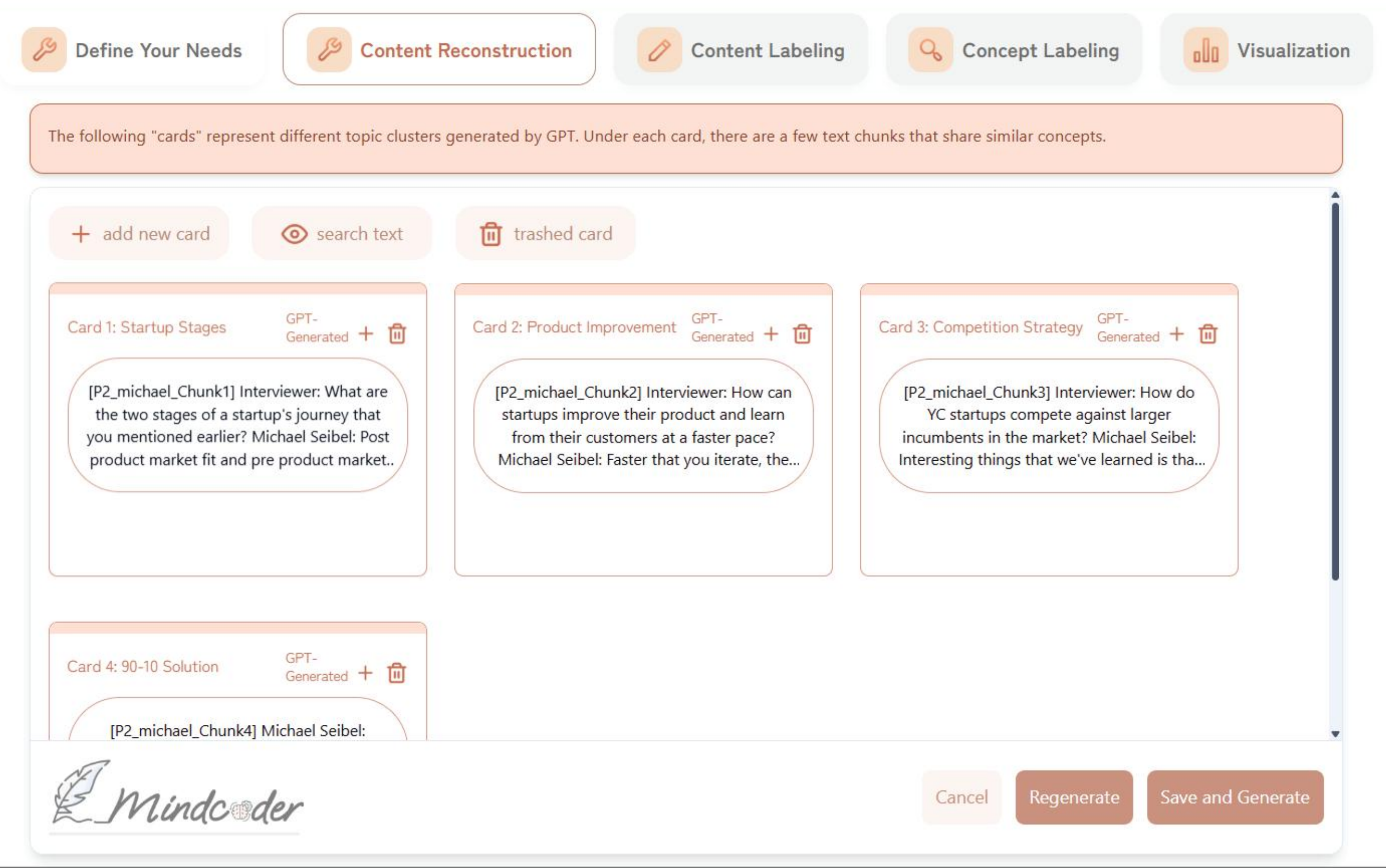}
        \caption{Card}
        \label{fig:card}
    \end{subfigure}

    \begin{subfigure}[t]{0.33\linewidth}
        \centering
        \includegraphics[width=\linewidth]{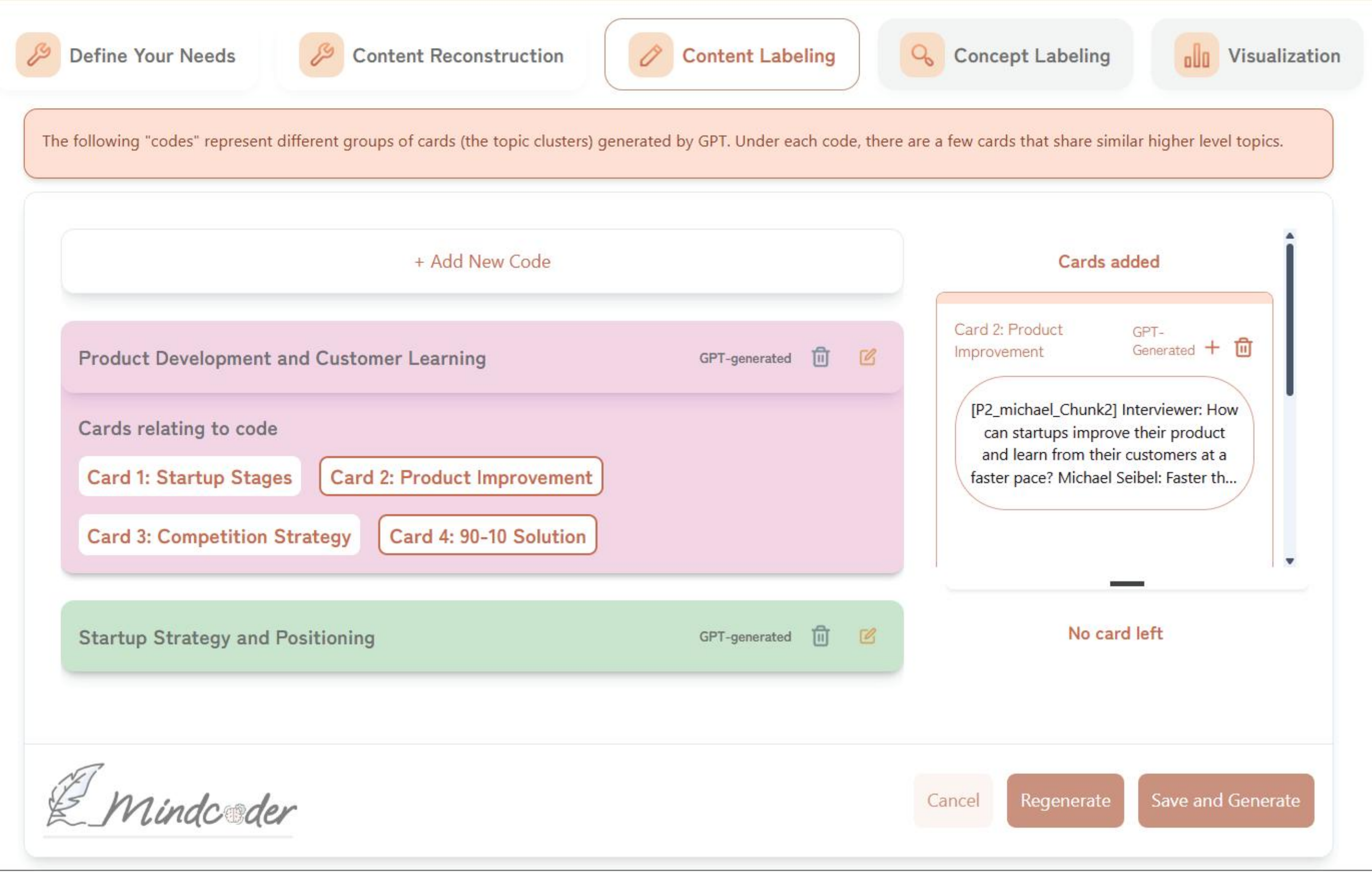}
        \caption{Code}
        \label{fig:code}
    \end{subfigure}
    \begin{subfigure}[t]{0.33\linewidth}
        \centering
        \includegraphics[width=\linewidth]{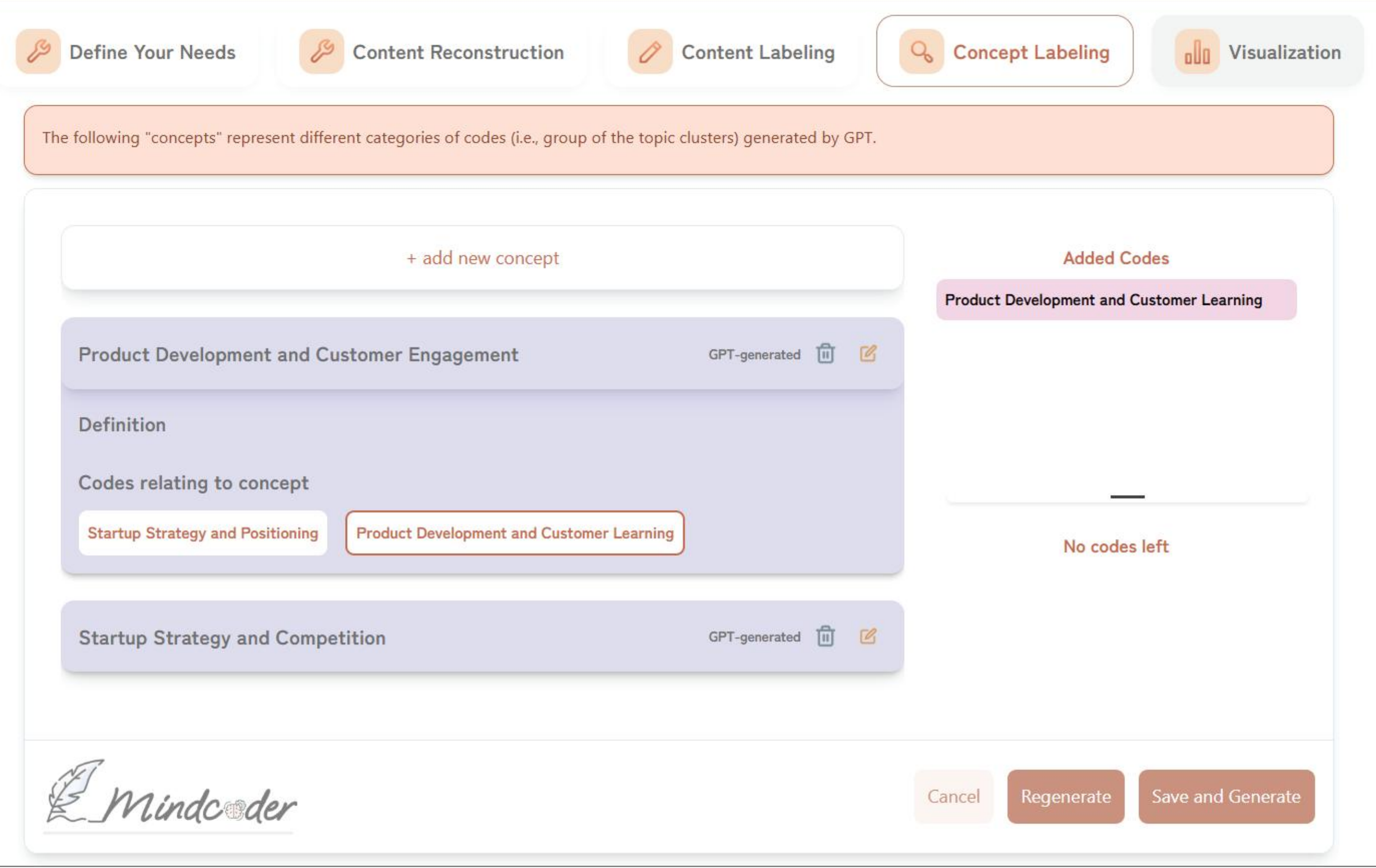}
        \caption{Concept}
        \label{fig:concept}
    \end{subfigure}
    \begin{subfigure}[t]{0.33\linewidth}
        \centering
        \includegraphics[width=\linewidth]{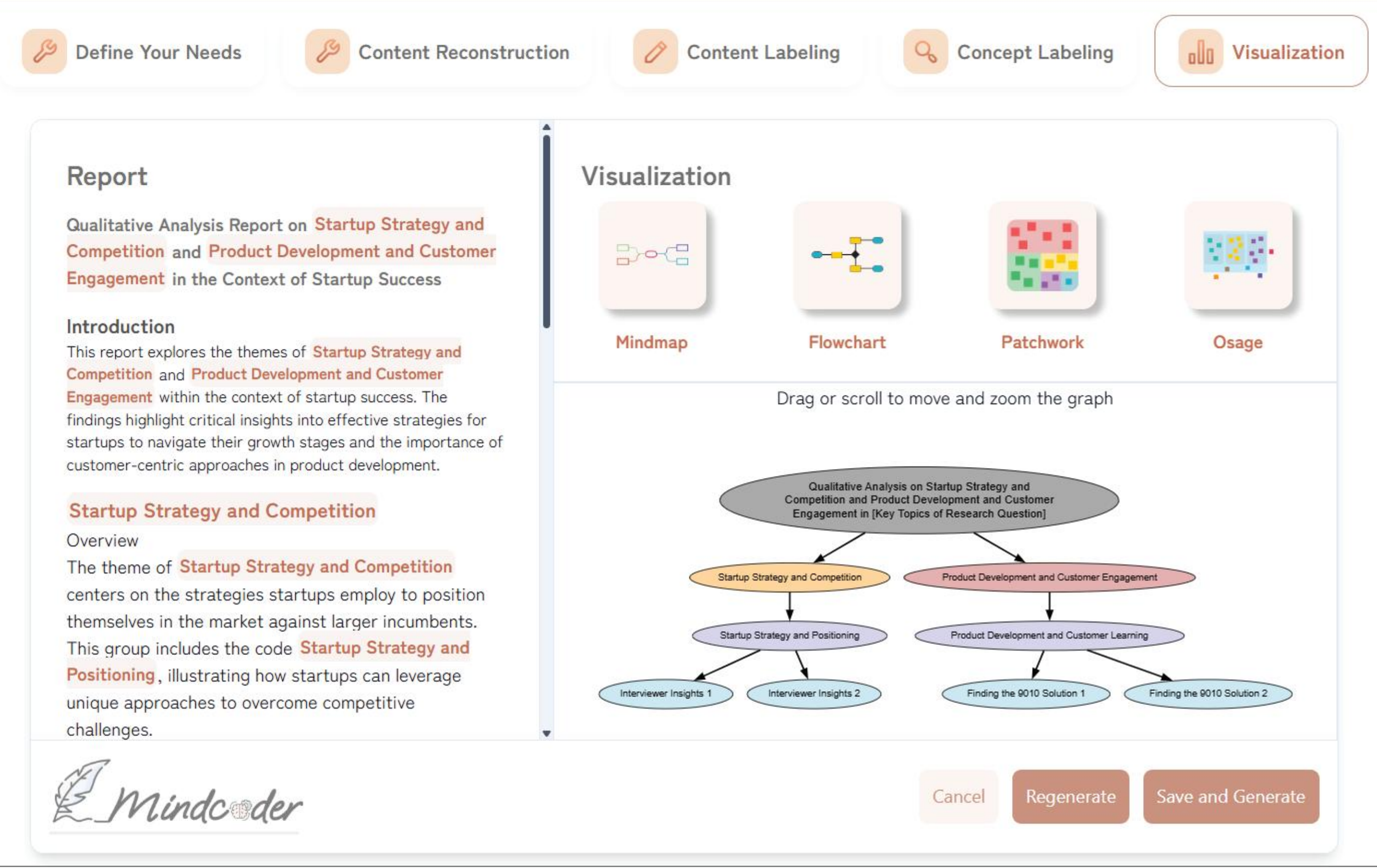}
        \caption{Graph}
        \label{fig:graph}
    \end{subfigure}

    \caption{Design Probe Used in Formative Interview}
    \label{fig:mindcoder-v1}
\end{figure*}

\section{Final MindCoder Interface}

\begin{figure*}[!htbp]
    \centering
    \begin{subfigure}[!t]{0.48\linewidth}
        \centering
        \includegraphics[width=\linewidth]{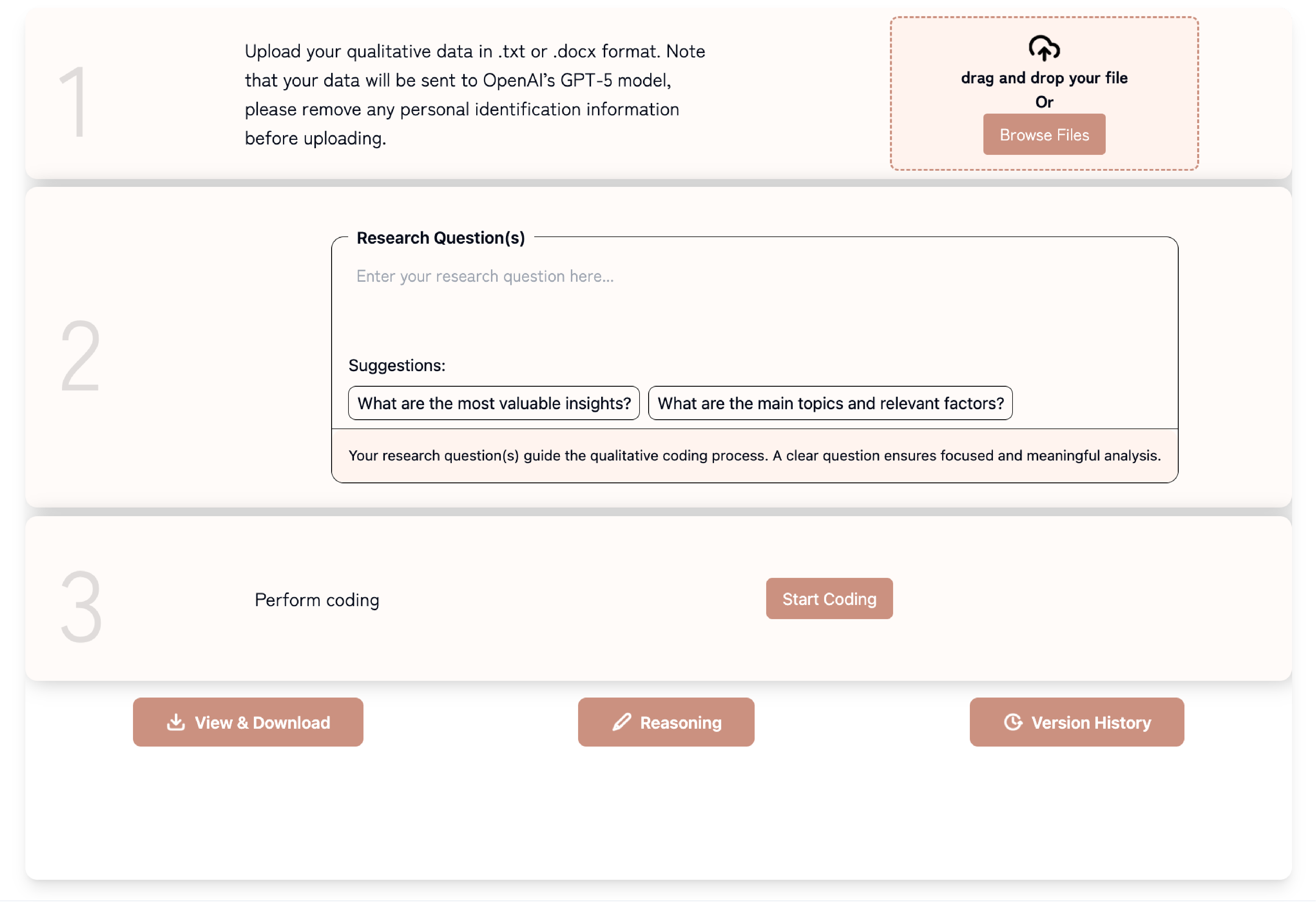}
        \caption{Data uploading and Research Questions}
        \label{fig:final-rqs}
    \end{subfigure}
    \begin{subfigure}[!t]{0.48\linewidth}
        \centering
        \includegraphics[width=\linewidth]{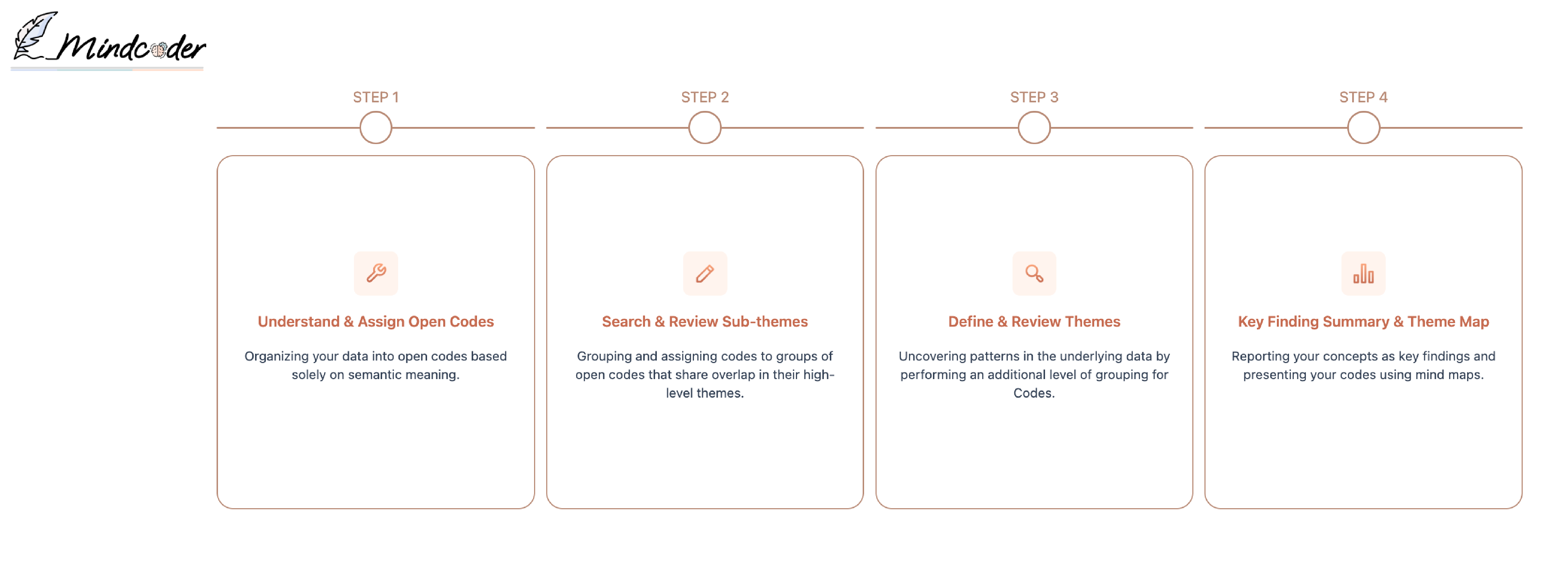}
        \caption{QDA steps}
        \label{fig:final-steps}
    \end{subfigure}
    \\
    \begin{subfigure}[!b]{0.48\linewidth}
        \centering
        \includegraphics[width=\linewidth]{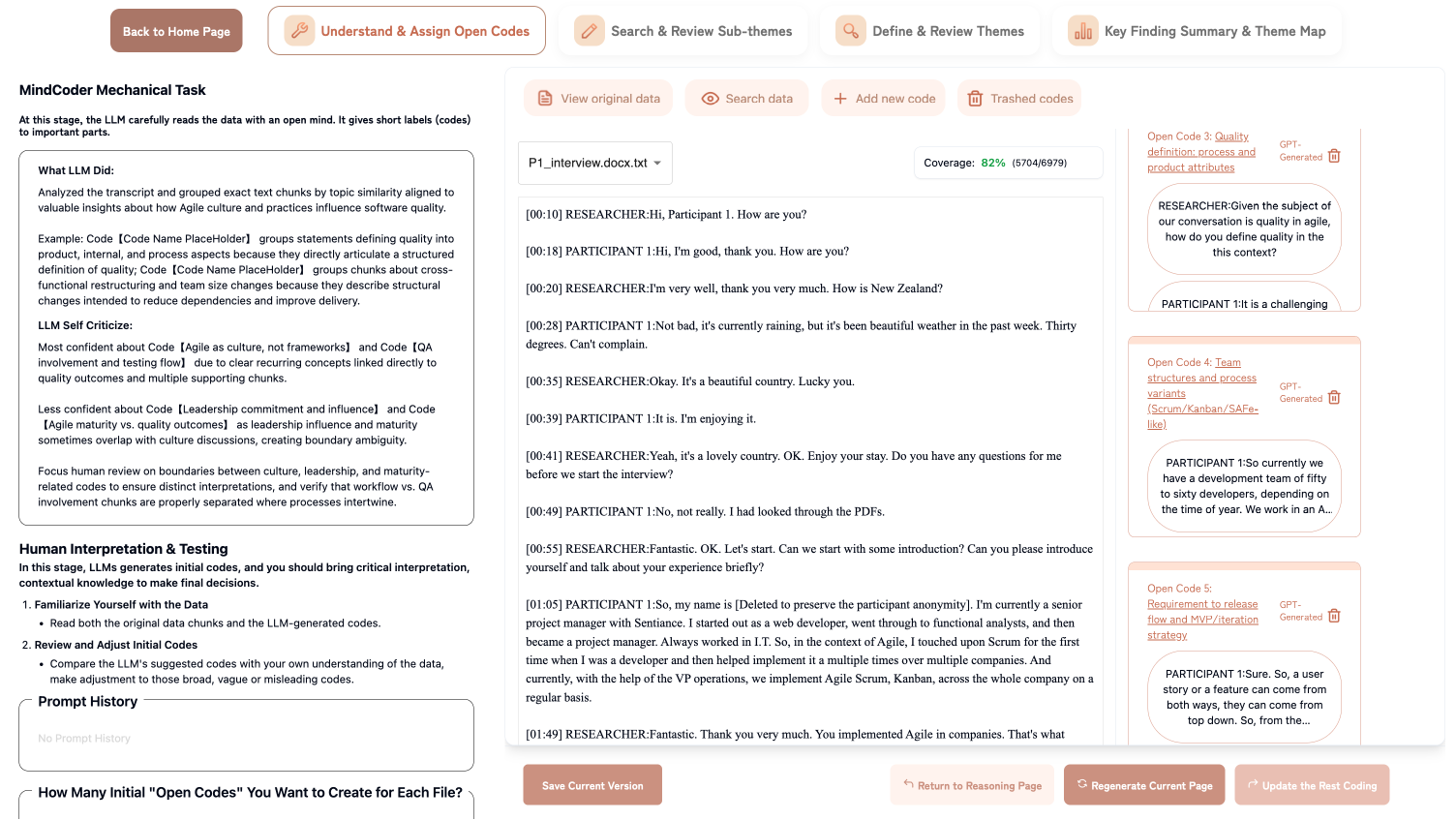}
        \caption{View Data}
        \label{fig:view data}
    \end{subfigure}
        \begin{subfigure}[!b]{0.48\linewidth}
        \centering
        \includegraphics[width=\linewidth]{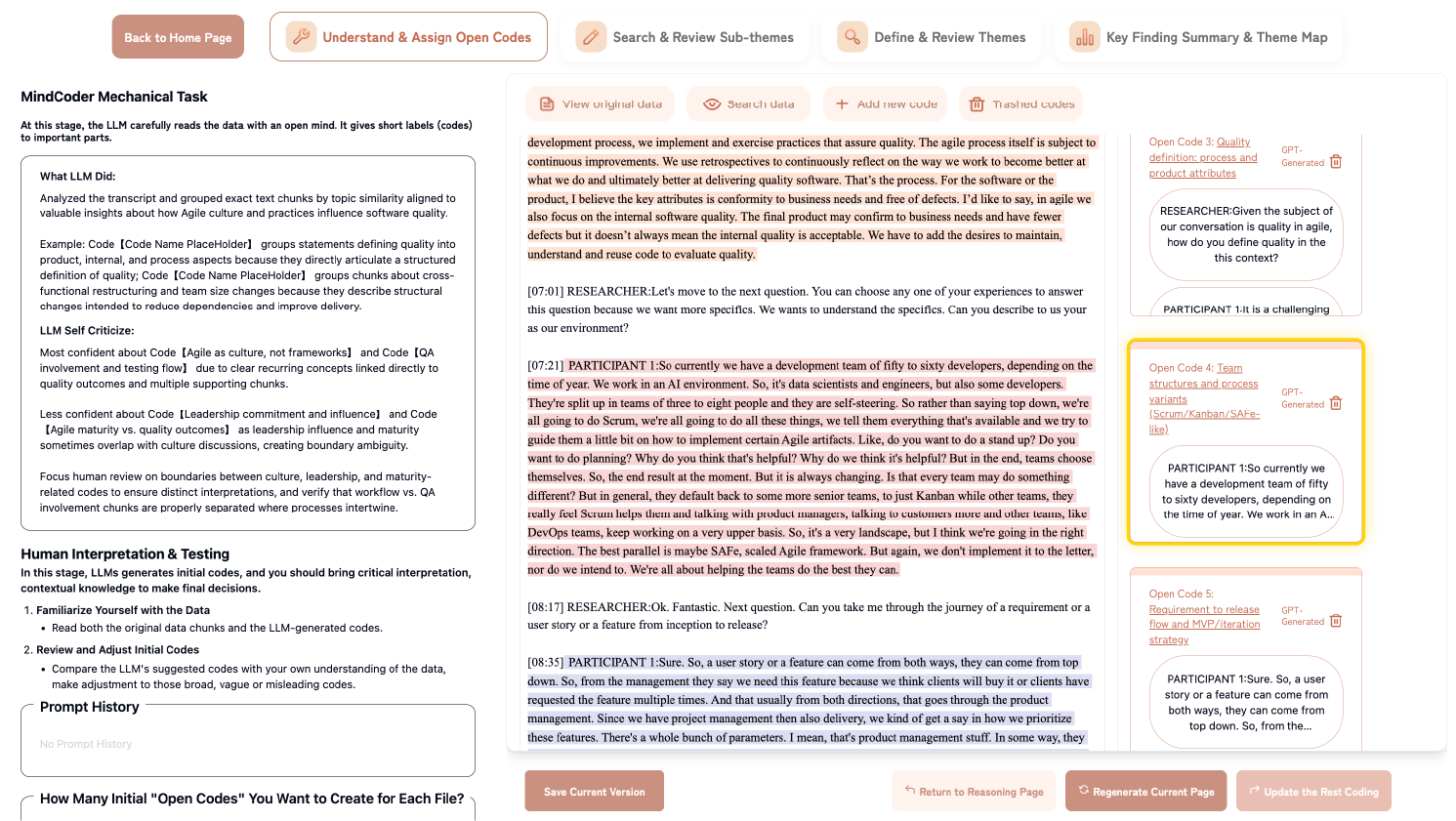}
        \caption{View Data2}
        \label{fig:view data2}
    \end{subfigure}

    \caption{Part of the Final Interface}
    \label{fig:final-mindcoder}
\end{figure*}

\section{Prompt list}
\label{sec:prompt_list}
Apart from following best practices of prompt design (e.g., structured input and output through JSON format, inclusion of few-shot examples), we also explicitly instructed the LLM to follow strict requirements such as \textit{“Do not alter, paraphrase, or revise any part of the original contents. Each chunk must contain the EXACT SAME text as it appears in the original data.”} The design of the final prompt went through four rounds of iteration, during which we refined the requirements, task descriptions, output format, and other identified issues. In the end, the final prompt contained around 6000 words. All prompt is attached in supplementary materials.


\end{document}